\documentclass[11pt,a4paper]{article}
\usepackage[cp1251]{inputenc} 
\usepackage{amsmath,amsthm,amssymb,epsfig,latexsym}
\usepackage{ulem}
\usepackage[numbers,square,sort&compress]{natbib}
\usepackage[english]{babel}
\usepackage{color}

\newtheorem{prop}{Proposition}[section]
\newtheorem{lemma}{Lemma}[section]

 \makeatletter
 \@addtoreset{equation}{section}
 \makeatother

\def\rvec{|0\>}

\def\beq#1{\begin{equation}\label{#1}}
\def\ba#1{\begin{multline}\label{#1}}
\def\eeq{\end{equation}}
\def\ea{\end{multiline}}

\def\Izer{{\sf K}}
\def\fun{{\sf f}}
\def\gun{{\sf g}}

\def\<{\langle}
\def\>{\rangle}
\def\CC{{\mathbb C}}

\def\r#1{(\ref{#1})}

\def\ot{\otimes}

\def\bbb{\mathbb{B}}

\def\RR{{\rm R}}

\def\qed{\hfill$\square$\medskip}

\def\blac{\bar u^{\scriptscriptstyle C}}
\def\blab{\bar u^{\scriptscriptstyle B}}
\def\bmuc{\bar v^{\scriptscriptstyle C}}
\def\bmub{\bar v^{\scriptscriptstyle B}}
\newcommand{\bla}{\bar u}

\def\E{{\sf E}}

\def\ccc{\mathbb{C}}

\def\muu{v}
\newcommand{\so}{{\scriptscriptstyle \rm I}}
\newcommand{\st}{{\scriptscriptstyle \rm I\hspace{-1pt}I}}
\newcommand{\sth}{{\scriptscriptstyle \rm I\hspace{-1pt}I\hspace{-1pt}I}}

\def\gone{{\sf g}}
\def\gtwo{{\sf g}}
\def\Uq#1{U_q(\widehat{\mathfrak{gl}}_{#1})}

\def\rrr{{\sf r}}
\def\bu{\bar u}
\def\bv{\bar v}
\def\bw{\bar w}

\newcommand{\num}{\\\rule{0pt}{20pt}}

\newcommand{\Izerl}{\Izer^{(l)}}
\newcommand{\Izerr}{\Izer^{(r)}}
\newcommand{\Izerlr}{\Izer^{(l,r)}}
\newcommand{\Izerrl}{\Izer^{(r,l)}}
\newcommand{\RHCl}{{\sf Z}^{(l)}}
\newcommand{\RHCr}{{\sf Z}^{(r)}}
\newcommand{\RHClr}{{\sf Z}^{(l,r)}}
\newcommand{\RHCrl}{{\sf Z}^{(r,l)}}
\begin{document}

\renewcommand*{\thefootnote}{\fnsymbol{footnote}}
\begin{flushright}
LAPTH-065/13
\end{flushright}

\vspace{20pt}

\begin{center}
\begin{LARGE}
{\bf Scalar products in  models with $GL(3)$ \\[1.2ex]
 trigonometric $R$-matrix. Highest coefficient}
\end{LARGE}

\vspace{40pt}

\begin{large}
{S.~Pakuliak${}^a$, E.~Ragoucy${}^b$, N.~A.~Slavnov${}^c$\footnote{pakuliak@theor.jinr.ru, eric.ragoucy@lapth.cnrs.fr, nslavnov@mi.ras.ru}}
\end{large}

 \vspace{12mm}

${}^a$ {\it Laboratory of Theoretical Physics, JINR, 141980 Dubna, Moscow reg., Russia,\\
Moscow Institute of Physics and Technology, 141700, Dolgoprudny, Moscow reg., Russia,\\
Institute of Theoretical and Experimental Physics, 117259 Moscow, Russia}

\vspace{4mm}

${}^b$ {\it Laboratoire de Physique Th\'eorique LAPTH, CNRS and Universit\'e de Savoie,\\
BP 110, 74941 Annecy-le-Vieux Cedex, France}

\vspace{4mm}

${}^c$ {\it Steklov Mathematical Institute,
Moscow, Russia}

\vspace{4mm}


\end{center}

\vspace{2mm}

\begin{abstract}
We study quantum integrable models with
$GL(3)$ trigonometric $\RR$-matrix solvable by the nested algebraic Bethe ansatz.
Scalar products of Bethe vectors in such models can be expressed in terms of a bilinear combination of the
highest coefficients.  We show that
in the models  with $GL(3)$ trigonometric $\RR$-matrix there exist two different highest coefficients. We obtain various  representations for them in terms of sums over partitions. We also prove several important properties of the highest coefficients, which are necessary for
the evaluation of the scalar products.
\end{abstract}

\vspace{4mm}



\renewcommand*{\thefootnote}{\arabic{footnote}}
\addtocounter{footnote}{-1}

\section{Introduction}

One of the striking facts about quantum integrable systems is the possibility to find the Hamiltonian eigenvectors.  Then, the knowledge of these eigenvectors  allows one to have analytical insight on the form and the behavior of correlation functions for these models. The general framework for such calculation is the Quantum Inverse Scattering Method
\cite{FadST79,KulRes83,BogIK93L,FadLH96}, and the use of the Bethe ansatz to construct the eigenvectors of the transfer matrix,
which is a generating functional of all commuting Hamiltonians. Unfortunately, if the method works well for the ``simplest'' cases based on $GL(2)$ or $\Uq{2}$ symmetries, it becomes quickly very technical for models based on algebras of higher rank, and much less is known in these later cases.

In the present paper we begin a systematic study of scalar products of the Bethe vectors in quantum integrable models with
$GL(3)$ trigonometric $\RR$-matrix. The role of the scalar products is extremely important in the study
of correlation functions \cite{Kor82,IzeKor84,Kor84,BogIK93L}. In particular, focusing on the class of  quantum integrable models where the inverse scattering problem can be solved \cite{KitMaiT99,MaiTer00}, one can reduce the problem of calculation of the form factors and the
correlation functions of  local operators to the calculation of the scalar products of the Bethe vectors  \cite{KitMaiT99}.  Furthermore explicit and compact formulas for the scalar products sometimes allow one
to study  the correlation functions even in such models, for which the solution of the inverse scattering problem is not known
\cite{Kor82,IzeKor84,Kor84,KitKMST07,BogIK93L}.  This approach was successfully applied for the quantum integrable models with $GL(2)$-invariant or $GL(2)$  trigonometric $\RR$-matrix
\cite{KitMT00,KitMST02,KitKMST09b,GohKS04,GohKS05,SeeBGK07,KitKMST11,KitKMST12,CauHM05,PerSCHMWA06,PerSCHMWA07,CauCS07}.
In all these works a determinant representation for the scalar products of the Bethe vectors obtained in \cite{Sl} was essentially used.

The problem of the scalar products appears to be much more sophisticated in the models based on the higher rank algebras.
The first results in this field  were obtained by N.~Reshetikhin  for the models with $GL(3)$-invariant
$\RR$-matrix \cite{Res86}. There, a
formula  for the scalar product of generic Bethe vectors  and a determinant representation
for the norm of the transfer matrix eigenvectors were found.
In the Reshetikhin representation for the scalar product, the notion of ``highest coefficient'' plays the most important role.
This function depends on the $\RR$-matrix of the model
and appears to be a rational function of the Bethe parameters.
The scalar product is a bilinear combination of these highest coefficients. The knowledge of the highest coefficient
allows one, in some important particular cases, to reduce this bilinear combination to a determinant
representation \cite{Whe12,BelPRS12b,BelPRS13a} analogous to the one of \cite{Sl}.

It was shown in \cite{Res86}  that the highest coefficient is equal to a partition function of the $15$-vertex
model with special boundary conditions. Using this fact one can obtain explicit representations
for the highest coefficient in models with the $GL(3)$-invariant $\RR$-matrix \cite{Whe12,BelPRS12a}. Unfortunately, these results can not
be directly extended to the case of models with $GL(3)$ trigonometric  $\RR$-matrix. The main reason
is that the $GL(3)$ trigonometric  $\RR$-matrix is not symmetric (see \eqref{UqglN-R}). This leads
to the fact that in these models  actually there are two highest
coefficients, which have essentially different explicit representations.
The main purpose of this paper is to derive these explicit formulas. We also establish a number of important properties of the highest coefficients, which are necessary for the calculation of the scalar products of the Bethe vectors.

In contrast to the Reshitihin's approach, we do not associate  the highest coefficients with some partition functions.
Instead we use a more direct method for their calculation.  The first tool of our approach is  an explicit representation for the dual Bethe vectors \cite{BelPRS13c}. It is worth mentioning that in pioneer papers on the nested Bethe ansatz \cite{KulRes83,KulRes81,KulRes82} no explicit formulas for the Bethe vectors and the dual ones were given.
More detailed formulas were obtained in \cite{VT} in the theory of solutions of the
quantum Knizhnik--Zamolodchikov equation. There the Bethe vectors were given by certain trace over
auxiliary spaces of the products of  monodromy matrices and $\RR$-matrices.

Explicit expressions for the Bethe vectors  in terms of the monodromy  matrix elements
for the models with the $GL(N)$ trigonometric $\RR$-matrix were obtained in the work
\cite{KP-GLN}, where  the realization of Bethe vectors in terms of the current generators of the quantum affine algebra $\Uq{N}$
\cite{EKhP} was used (see also \cite{OPS}).

The second tool of our method is based on the formulas of the multiple action of
the monodromy matrix entries onto the Bethe vectors \cite{PakRS13a}. Using these formulas one
can calculate not only the highest coefficients, but the whole scalar product of the Bethe vectors.
However, the last problem is much more technical. It requires, in particular, the knowledge of several
non-obvious properties of the highest coefficients. Therefore we postpone its solution to our further
publication. In the present paper we restrict ourselves with the study of the highest coefficients only.

The plan of the paper is as follows. In section~\ref{sec:bckgr}, we present the model we work with, and introduce the notations that will be used throughout the paper. We also recall some results obtained previously and needed here. In section~\ref{sec:sumHC}, we exhibit the main result of the paper, a sum formulas for the highest coefficients. The proof of the sum formulas is given in
section~\ref{S-SerHC}. Section~\ref{sec:altern} gathers different properties of the highest coefficients, as well as some alternative presentations for them. Appendices collect different formulas or proofs of formulas, needed in the paper.

\section{General background\label{sec:bckgr}}

\subsection{The model}
We consider a quantum integrable model defined by the monodromy matrix $T(u)$ with the matrix elements $T_{ij}(u)$,
$i,j=1,2,3$ which satisfies the commutation relation
\begin{equation}\label{RTT}
\RR(u,v)\cdot (T(u)\ot \mathbf{1})\cdot (\mathbf{1}\ot T(v))=
(\mathbf{1}\ot T(v))\cdot (T(u)\ot \mathbf{1})\cdot \RR(u,v),
\end{equation}
with the $GL(3)$ trigonometric quantum $\RR$-matrix
\begin{equation}\label{UqglN-R}
\begin{split}
\RR(u,v)\ =\ \fun(u,v)&\ \sum_{1\leq i\leq 3}\E_{ii}\ot \E_{ii}\ +\
\sum_{1\leq i<j\leq 3}(\E_{ii}\ot \E_{jj}+\E_{jj}\ot \E_{ii})
\\
+\ &\sum_{1\leq i<j\leq 3}
(u\gone(u,v) \E_{ij}\ot \E_{ji}+ v\gtwo(u,v)\E_{ji}\ot \E_{ij})\,.
\end{split}
\end{equation}
Here the rational functions $\fun(u,v)$ and $\gone(u,v)$   are
\begin{equation}\label{fgg}
\fun(u,v)=\frac{qu-q^{-1}v}{u-v},\quad \gone(u,v)=\frac{q-q^{-1}}{u-v}\,,
\end{equation}
where $q$ is a complex number (a deformation parameter), and $(\E_{ij})_{lk}=\delta_{il}\delta_{jk}$, $i,j,l,k=1,2,3$ are $3\times3$ matrices
with unit in the intersection of $i$th row and $j$th column and zero matrix elements elsewhere.
The $\RR$-matrix \r{UqglN-R} is called ``trigonometric'' because its classical limit gives the classical trigonometric $r$-matrix \cite{BelDri82}. The trigonometric $\RR$-matrix \r{UqglN-R} is written in multiplicative variables and depends actually on the ratio $u/v$ of these multiplicative parameters.

Due to the commutation relation \r{RTT} the transfer matrix $t(w)={\rm tr}\ T(w)=T_{11}(w)+T_{22}(w)+T_{33}(w)$
generates a set of commuting integrals of motion and the first step of the algebraic Bethe ansatz \cite{FadST79,KulRes83} is the construction
of the set of eigenstates for these commuting operators in terms  of the monodromy
matrix entries. We assume that these matrix elements act in a quantum space $V$ and this
space possesses a vector $\rvec\in V$ such that
\begin{equation}\label{rsba}
T_{ij}(u)\rvec =0,\quad i>j,\quad T_{ii}(u)\rvec= \lambda_i(u)\rvec\,,\qquad \lambda_i(u)\in\CC[[u,u^{-1}]]\,.
\end{equation}
We also assume that the operators $T_{ij}(u)$ act in a dual space $V^*$ with
a vector $\langle 0|\in V^*$ such that
\begin{equation}\label{drsba}
\langle 0|T_{ij}(u) =0,\quad i<j,\quad  \langle 0|T_{ii}(u)= \lambda_i(u)\langle 0|\,,
\end{equation}
and  $\lambda_i$ are the same as in \eqref{rsba}.

The Bethe vectors ${\bbb}^{a,b}(\bu;\bv)$
in quantum integrable models with a
$GL(3)$ trigonometric $\RR$-matrix depend on two sets
of  variables
\begin{equation}\label{set111}
\bu = \left\{u_1,\ldots,u_a\right\},\quad \bv=\left\{v_1,\ldots,v_b\right\}\,,
\end{equation}
which are called the Bethe parameters. These vectors can be constructed in the framework of the nested Bethe
ansatz method formulated in \cite{KulRes83} and are given by certain polynomials in the monodromy matrix elements
$T_{12}(u)$, $T_{23}(u)$, $T_{13}(u)$ depending on the Bethe parameters and applied to the vector $|0\rangle$.
They become eigenstates of the transfer matrix $t(w)$, if the parameters $\bu$ and $\bv$ satisfy the system of
Bethe equations \cite{KulRes83}. Such vectors sometimes are called {\sl on-shell} Bethe vectors. Otherwise,
if $\bu$ and $\bv$ are generic complex numbers, we deal with generic Bethe vectors.

Similarly, dual vectors ${\ccc}^{a,b}(\bu;\bv)$ can be constructed as  polynomials in
$T_{21}(u)$, $T_{32}(u)$, $T_{31}(u)$  applied to the vector $\langle0|$. They also depend on two
sets of Bethe parameters $\bu$ and $\bv$ (see \eqref{Or-form}, \eqref{Or-form2} for the explicit formulas)
and become eigenstates of the transfer matrix $t(w)$, if the  sets of parameters $\bu$ and $\bv$ satisfy the system of
Bethe equations.

\subsection{Notations}

Below we always denote sets of variables by bar, like in \eqref{set111}. If a set of variables is
multiplied by a number $\alpha\bu$ (in particular, $\bu q^{\pm 2}$), then it means that all the
elements of the set are multiplied by this number
\begin{equation}\label{set-numb}
\alpha\bu = \left\{\alpha u_1,\ldots,\alpha u_a\right\},\qquad
\bv q^{\pm 2}=\left\{v_1q^{\pm 2},\ldots,v_bq^{\pm 2}\right\}\,.
\end{equation}

To save space and simplify the presentation,  we  use the following convention for the products of the
commuting entries of the monodromy matrix $T_{ij}(w)$, the vacuum eigenvalues $\lambda_i(w)$ and their ratios
$\rrr_k(w)=\lambda_k(w)/\lambda_2(w)$, $k=1,3$. Namely, whenever such an operator or a scalar function depends on a set of variables (for instance, $T_{ij}(\bw)$, $\lambda_i(\bu)$, $\rrr_k(\bv)$),   this means that we
deal with the product of the operators or the scalar functions with respect to the corresponding set:
 \begin{equation}\label{SH-prod}
 T_{ij}(\bar w)=\prod_{w_k\in\bar w}   T_{ij}(w_k);\quad
  \lambda_2(\bar u)=\prod_{u_j\in\bar u}  \lambda_2(u_j);\quad
 \rrr_k(\bar v_\ell)= \prod_{\substack{v_j\in\bar v\\v_j\ne v_\ell}} \rrr_k(v_j).
 \end{equation}
 Here and below the notation $\bv_\ell$ for an
arbitrary set $\bv$ means the set $\bv\setminus v_\ell$.
A similar convention will be used for the products of functions $\fun(u,v)$
\begin{equation}\label{SH-prod1}
 \fun(w_i, \bar w_i)= \prod_{\substack{w_j\in\bar w\\w_j\ne w_i}} \fun(w_i, w_j);\quad
 \fun(\bar u,\bar v)=\prod_{u_j\in\bar u}\prod_{v_k\in\bar v} \fun(u_j,v_k).
 \end{equation}
Partitions of sets into two or more subsets will be noted as $\bar u \Rightarrow \{\bar u_{\so}\,,\, \bar u_{\st}\}$. Here the roman numbers are used for the numeration of subsets
$\bar u_{\so}$ and  $\bar u_{\st}$. Union of sets is denoted by braces, for example, $\{\bw,\bu\}=\bar\eta$.

In various formulas the Izergin determinant $\Izer_k(\bar x|\bar y)$ appears
\cite{Ize87}. It is defined
for two sets $\bar x$ and $\bar y$ of the same cardinality $\#\bar x=\#\bar y=k$:
\begin{equation}\label{Izer}
\Izer_k(\bar x|\bar y)=\frac{\prod_{1\leq i,j\leq k}(qx_i-q^{-1}y_j)}
{\prod_{1\leq i<j\leq k}(x_i-x_j)(y_j-y_i)}
\cdot\det \left[\frac{q-q^{-1}}{(x_i-y_j)(qx_i-q^{-1}y_j)}\right]\,.
\end{equation}
Below we also use two modifications of the Izergin determinant
\begin{equation}\label{Mod-Izer}
\Izerl_k(\bar x|\bar y)= \prod_{i=1}^kx_i\cdot\Izer_k(\bar x|\bar y)\,, \qquad
\Izerr_k(\bar x|\bar y)= \prod_{i=1}^ky_i\cdot\Izer_k(\bar x|\bar y)\,,
\end{equation}
which we call left and right Izergin determinants respectively.
Some properties of the Izergin determinant and its modifications are gathered in Appendix~\ref{A-PID}.

The left and the right Izergin determinants play the role of the highest coefficients of scalar products in
the models based on the $\Uq{2}$ algebra. Certainly the difference between them is very small, in particular,
\begin{equation}\label{Diff-Mod-Izer}
\prod_{i=1}^kx_i^{-1}\cdot\Izerl_k(\bar x|\bar y)-
\prod_{i=1}^ky_i^{-1}\cdot\Izerr_k(\bar x|\bar y)= 0\,.
\end{equation}
Moreover in the $\Uq{2}$ algebra there exists a  transformation that makes the $\RR$-matrix symmetric. This
map, being applied to the scalar products, makes the two highest coefficients equal to each other
(in this case they both are given as the original Izergin determinant $\Izer_k$ multiplied by the product of
$\sqrt{x_iy_i}$). However in the models based on the higher rank algebras the mentioned  transformation of the
$\RR$-matrix no longer exists. This leads to the fact that the analogs of the right and left highest coefficients in these models have essentially different representations.

\subsection{Multiple action of  the operators $T_{ij}$ on Bethe vectors}

For the derivation of explicit representations of the highest coefficients we need to know
the  actions of products $T_{ij}(\bar w)$ onto the Bethe vectors. They have been computed in \cite{PakRS13a}.
We recall here the ones that we use in this paper.
 Below everywhere $\{\bar\muu,\bar w\}=\bar\xi$, $\{\bla,\bar w\}=\bar\eta$ and $\#\bar w=n$.

The multiple action of $T_{21}$ is given by
 \begin{multline}\label{act21-1}
 T_{21}(\bar w)\mathbb{B}^{a,b}(\bar u;\bar v)=(-q)^n\lambda_2(\bar w)\,\sum
\rrr_1(\bar\eta_{\so})\;
\fun(\bar\eta_{\st},\bar\eta_{\so})\fun(\bar\eta_{\st},\bar\eta_{\sth})\fun(\bar\eta_{\sth},\bar\eta_{\so})
 \frac{\fun(\bar\xi_{\st},\bar\xi_{\so})}{\fun(\bar\xi_{\st},\bar\eta_{\so})} \\
 \times  \Izerr_n(q^{-2}\bar w|\bar\eta_{\st})
  \Izerl_n(\bar\eta_{\so}|q^2\bar\xi_{\so})\Izerl_n(\bar\xi_{\so}|q^2\bar w)
 \,\mathbb{B}^{a-n,b}(\bar\eta_{\sth};\bar\xi_{\st}).
 \end{multline}
The sum is taken over partitions of: $\bar\eta\Rightarrow\{\bar\eta_{\so},\bar\eta_{\st},\bar\eta_{\sth}\}$ with $\#\bar\eta_{\so}=\#\bar\eta_{\st}=n$; and
  $\bar\xi\Rightarrow\{\bar\xi_{\so},\bar\xi_{\st}\}$
 with $\#\bar\xi_{\so}=n$.

The multiple action of $T_{32}$ is given by
  \begin{multline}\label{act32-1}
 T_{32}(\bar w)\mathbb{B}^{a,b}(\bar u;\bar v)=
 (-q)^{-n}\lambda_2(\bar w)\,\sum \rrr_3(\bar\xi_{\so})\;
\fun(\bar\xi_{\so},\bar\xi_{\st})\fun(\bar\xi_{\so},\bar\xi_{\sth})\fun(\bar\xi_{\sth},\bar\xi_{\st})
 \frac{\fun(\bar\eta_{\so},\bar\eta_{\st})}{\fun(\bar\xi_{\so},\bar\eta_{\st})} \\
 \times  \Izerr_n(q^{-2}\bar w|\bar\eta_{\so})
 \Izerr_n(q^{-2}\bar\eta_{\so}|\bar\xi_{\so})\Izerl_n(\bar\xi_{\st}|q^2\bar w)
 \,\mathbb{B}^{a,b-n}(\bar\eta_{\st};\bar\xi_{\sth}).
 \end{multline}
The sum is taken over partitions of: $\bar\xi\Rightarrow\{\bar\xi_{\so},\bar\xi_{\st},\bar\xi_{\sth}\}$ with $\#\bar\xi_{\so}=\#\bar\xi_{\st}=n$; and
  $\bar\eta\Rightarrow\{\bar\eta_{\so},\bar\eta_{\st}\}$
 with $\#\bar\eta_{\so}=n$.

{\sl Remark.} Note that the restrictions on the cardinalities of subsets in the formulas \eqref{act21-1} and \eqref{act32-1} are shown explicitly by the subscripts of the Izergin determinants and the superscripts of the Bethe vectors. However, for convenience we will describe such the restrictions in special comments after formulas.

If we set  $n=a$ in \eqref{act21-1}, then $\bar\eta_{\sth}=\emptyset$,  and we obtain
 \begin{multline}\label{A-act21-2}
 T_{21}(\bar w)\mathbb{B}^{a,b}(\bar u;\bar v)=(-q)^a\lambda_2(\bar w)\sum
\rrr_1(\bar\eta_{\so})\;
 \frac{\fun(\bar\eta_{\st},\bar\eta_{\so})\fun(\bar\xi_{\st},\bar\xi_{\so})}{\fun(\bar\xi_{\st},\bar\eta_{\so})} \\
 \times  \Izerr_a(q^{-2}\bar w|\bar\eta_{\st})
  \Izerl_a(\bar\eta_{\so}|q^2\bar\xi_{\so})\Izerl_a(\bar\xi_{\so}|q^2\bar w)
 \,\mathbb{B}^{0,b}(\emptyset;\bar\xi_{\st}).
 \end{multline}

If in addition $\bar v=\emptyset$ and we want to find a coefficient of $\rrr_1(\bar w)$, then  $\bar\xi_{\so}=\bar w$, $\bar\xi_{\st}=\emptyset$, and we should set $\bar\eta_{\so}=\bar w$, $\bar\eta_{\st}=\bar u$. Using \eqref{K-red}
and \eqref{K-invers} we obtain
 \begin{equation}\label{A-act21-3}
 T_{21}(\bar w)\mathbb{B}^{a,0}(\bar u;\emptyset)=\lambda_2(\bar w)\rrr_1(\bar w)\Izerl_a(\bar u|\bar w)|0\rangle
 +\text{IT}\;,
\end{equation}
where IT stands for {\it irrelevant terms}, i.e. terms that do not contribute to the coefficient we consider.

Similarly, if we set $n=b$ in \eqref{act32-1}, then $\bar\xi_{\sth}=\emptyset$,  and we obtain
 \begin{multline}\label{A-act32-2}
 T_{32}(\bar w)\mathbb{B}^{a,b}(\bar u;\bar v)=
 (-q)^{-b}\lambda_2(\bar w)\sum \rrr_3(\bar\xi_{\so})\;
 \frac{\fun(\bar\xi_{\so},\bar\xi_{\st})\fun(\bar\eta_{\so},\bar\eta_{\st})}{\fun(\bar\xi_{\so},\bar\eta_{\st})} \\
 \times  \Izerr_b(q^{-2}\bar w|\bar\eta_{\so})
 \Izerr_b(q^{-2}\bar\eta_{\so}|\bar\xi_{\so})\Izerl_b(\bar\xi_{\st}|q^2\bar w)
 \,\mathbb{B}^{a,0}(\bar\eta_{\st};\emptyset).
 \end{multline}
If in addition $\bar u=\emptyset$ and we want to find a coefficient of $\rrr_3(\bar w)$, then  $\bar\eta_{\so}=\bar w$, $\bar\eta_{\st}=\emptyset$, and we should set $\bar\xi_{\so}=\bar w$, $\bar\xi_{\st}=\bar v$. Using \eqref{K-red}
and \eqref{K-invers} we obtain
 \begin{equation}\label{A-act32-3}
 T_{32}(\bar w)\mathbb{B}^{0,b}(\emptyset;\bar v)=\lambda_2(\bar w)\rrr_3(\bar w)\Izerr_b(\bar w|\bar v)|0\rangle
 +\text{IT}\;.
\end{equation}
Observe  that the actions \eqref{A-act21-3}, \eqref{A-act32-3} reproduce the known results
for the models with  $GL(2)$ trigonometric  $\RR$-matrix.

\subsection{Dual Bethe vectors}

We have mentioned already that the Bethe vectors are given by certain polynomials in the monodromy matrix elements
$T_{12}(u)$, $T_{23}(u)$, $T_{13}(u)$ applied to the vector $|0\rangle$. The explicit form of these
 polynomials is not essential in the formulas for the multiple action \eqref{act21-1}, \eqref{act32-1}.
 However we need explicit representations for the dual Bethe vectors in terms of the monodromy matrix elements
 in order to obtain formulas for the highest coefficients. Such representations were obtained in our  work
 \cite{BelPRS13c}. We give two of them:
%
\begin{equation}\label{Or-form}
\mathbb{C}^{a,b}(\bar u;\bar v) =\sum \frac{\Izerr_{k}(\bar v_{\so}|\bar u_{\so})}{\lambda_2(\bar v_{\st})\lambda_2(\bar u)}
\frac{\fun(\bar v_{\st},\bar v_{\so})\fun(\bar u_{\so},\bar u_{\st})}{\fun(\bar v,\bar u)}\,
\langle0|T_{32}(\bar v_{\st})T_{21}(\bar u_{\st})T_{31}(\bar u_{\so})\;,
\end{equation}
and
\begin{equation}\label{Or-form2}
\mathbb{C}^{a,b}(\bar u;\bar v) =\sum \frac{\Izerl_{k}({\bar v}_{\so}|{\bar u}_{\so})}{\lambda_2(\bar u_{\st})
\lambda_2(\bar v)}
\frac{\fun(\bar v_{\st},\bar v_{\so})\fun(\bar u_{\so},\bar u_{\st})}{\fun(\bar v,\bar u)}\,
\langle0|T_{21}({\bar u_{\st}})T_{32}({\bar v}_{\st}) T_{31}({\bar v}_{\so})\;.
\end{equation}
Here the sum goes over all partitions of the sets $\bu\Rightarrow\{\bu_{\so},\bu_{\st}\}$ and $\bv\Rightarrow\{\bv_{\so},\bv_{\st}\}$
such that $\#\bu_{\so}=\#\bv_{\so}=k$, $k=0,\dots,\min(a,b)$.

Both of  these representations are needed for our purpose. They correspond to two different embeddings of $\Uq{2}$ into $\Uq{3}$ algebra. It is also easy to check that \eqref{Or-form} and \eqref{Or-form2} are related by the isomorphism
$\varphi$ described in \cite{BelPRS13c}.  This isomorphism maps the original algebra $U_q(\widehat{\mathfrak{gl}}_3)$ to the algebra $U_{q^{-1}}(\widehat{\mathfrak{gl}}_3)$
\beq{phi}
\varphi\big( T_{i,j}(u)\big)\ =\tilde T_{4-j,4-i}(u)\,,
\eeq
where $T(u)\in\Uq{3}$ and $\tilde T(u)\in U_{q^{-1}}(\widehat{\mathfrak{gl}}_3)$ respectively. The map \eqref{phi}
is a very powerful tool for the study of the scalar products. In particular, many properties of the scalar products
can be established via the mapping between $\Uq{3}$ and $U_{q^{-1}}(\widehat{\mathfrak{gl}}_3)$.

\section{Sum formulas for the highest coefficients \label{sec:sumHC}}

The scalar products are defined as
\beq{Def-SP}
S_{a,b}(\blac;\bmuc|\blab;\bmub)=\ccc^{a,b}(\blac;\bmuc)\bbb^{a,b}(\blab;\bmub)\,,
\eeq
where all the Bethe parameters are generic complex numbers. We have added the superscripts $C$ and $B$
to the sets $\bar u$, $\bar v$ in order to stress that the vectors
$\ccc^{a,b}$ and $\bbb^{a,b}$ may depend on different sets of parameters.

Knowing the explicit form of the dual Bethe vectors \eqref{Or-form}, \eqref{Or-form2} and the multiple
action of the operators $T_{ij}$ \eqref{act21-1}, \eqref{act32-1} one can formally calculate
the scalar product\footnote{%
For the complete calculation of the scalar product one should also know the multiple action of the
operator $T_{31}(u)$. This action can be found in \cite{PakRS13a}. However for the calculation of
the highest coefficients this action is not needed.} \eqref{Def-SP}. It is clear that the
result is given as a sum with respect to partitions of the sets $\blac$, $\blab$, $\bmuc$, and $\bmub$.
The terms of this sum depend on the products of the vacuum eigenvalues $\rrr_1$ and $\rrr_3$, as well as
on the functions entering the $\RR$-matrix. In complete analogy with the case of $GL(3)$-invariant
$\RR$-matrix (see e.g. \cite{Res86}) one can derive the following representation:
\beq{scal}
S_{a,b}(\blac;\bmuc|\blab;\bmub)= \sum \frac{\rrr_1(\blac_\st)\rrr_1(\blab_\so)\rrr_3(\bmuc_\st)\rrr_3(\bmub_\so)}{\fun(\bmuc,\blac)\fun(\bmub,\blab)}\,
W_{\text{part}}\begin{pmatrix}\blac_{\st},\blab_{\st},&\blac_{\so},\blab_{\so}\\
\bmuc_{\so},\bmub_{\so},&\bmuc_{\st},\bmub_{\st}\end{pmatrix}.
\eeq
Here the sum runs over all the partitions $\blac\Rightarrow\{\blac_\so,\blac_\st\}$,
$\blab\Rightarrow\{\blab_\so,\blab_\st\}$,  $\bmuc\Rightarrow\{\bmuc_\so,\bmuc_\st\}$ and $\bmub\Rightarrow\{\bmub_\so,\bmub_\st\}$
with $\#\blac_{\so}=\#\blab_{\so}$ and $\#\bmuc_{\so}=\#\bmub_{\so}$.
The form of the functions $W_{\text{part}}$ depends on the partitions, what is shown by the subscript `$\text{part}$'. They also depend on  the $\RR$-matrix entries, but not on the functions $\rrr_1$ and $\rrr_3$. In other words, they depend on the algebra, not on the representations one chooses.

The highest coefficients $\RHClr_{a,b}$ are defined as  particular cases of the functions $W_{\text{part}}$, corresponding
to special choices of partitions:
\beq{def:Zlr}
\begin{aligned}
\RHCl_{a,b}(\blac;\blab|\bmuc;\bmub)& = W_{\text{part}}\begin{pmatrix}\blac,\blab,&\emptyset,\emptyset\\
\bmuc,\bmub,&\emptyset,\emptyset\end{pmatrix}\,,\\
\RHCr_{a,b}(\blab;\blac|\bmub;\bmuc)& =  W_{\text{part}}\begin{pmatrix}\emptyset,\emptyset,&\blac,\blab,\\
\emptyset,\emptyset,&\bmuc,\bmub\end{pmatrix}\,.
\end{aligned}
\eeq
Just as in the case of the Izergin determinant, we call these coefficients left and right. The subscripts of
the highest coefficients shows that $\#\blac=\#\blab=a$ and $\#\bmuc=\#\bmub=b$.

Similarly to the $GL(3)$-invariant case, all other coefficients
$W_{\text{part}}$ in \eqref{scal} can be expressed in term of products of the left and the right highest
coefficients\footnote{See \eqref{W-Reshet} for the explicit formula. The detailed proof  will be given in a forthcoming paper.}. Thus, the scalar product is a bilinear combination of the highest coefficients, and that is why the role of $\RHClr_{a,b}$ is so important.

In the case of the $GL(3)$-invariant $\RR$-matrix, the two highest coefficients coincide and are equal to a partition function
of the  $15$-vertex model with special boundary conditions \cite{Res86}. However, as we have already mentioned,
in the models with $GL(3)$ trigonometric $\RR$-matrix the left and right highest coefficients differ from
each other. The main result of the paper is an explicit expression for them.

\begin{prop}
The left and right highest coefficients have the following representations:
 \begin{equation}\label{RHC-IHC}
  \RHCl_{a,b}(\bar t;\bar x|\bar s;\bar y)=(-q)^{-b}\sum
 \Izerr_b(\bar s|\bar w_{\so}q^2)\Izerl_a(\bar w_{\st}|\bar t)
  \Izerl_b(\bar y|\bar w_{\so})\fun(\bar w_{\so},\bar w_{\st}),
 \end{equation}
 \begin{equation}\label{2-RHC-IHC}
  \RHCr_{a,b}(\bar t;\bar x|\bar s;\bar y)=(-q)^b\sum
 \Izerl_b(\bar s|\bar w_{\so}q^2)\Izerr_a(\bar w_{\st}|\bar t)
  \Izerr_b(\bar y|\bar w_{\so})\fun(\bar w_{\so},\bar w_{\st}).
 \end{equation}
Here $\bar w=\{\bar s,\;\bar x\}$. The sum is taken with respect to partitions of the set $\bar w\Rightarrow
\{\bar w_{\so},\bar w_{\st}\}$ with $\#\bar w_{\so}=b$ and $\#\bar w_{\st}=a$.
\end{prop}

We would like to draw the attention of the reader that the difference between
$\RHCl_{a,b}$ and $\RHCr_{a,b}$ is much more essential than the one
between $\Izerl_n$ and $\Izerr_n$. To see this one can consider the explicit expressions
for the highest coefficients in the simplest nontrivial case $a=b=1$:
 \begin{equation}\label{RHC-IHC-11}
 \begin{aligned}
  \RHCl_{1,1}(t;x|s;y)&=xy\; \gun(x,t)\gun(y,s)\fun(s,x)+xys \;\gun(x,s)\gun(s,t)\gun(y,x),\\
    \RHCr_{1,1}(t;x|s;y)&=ts\; \gun(x,t)\gun(y,s)\fun(s,x)+tsx \;\gun(x,s)\gun(s,t)\gun(y,x).
  \end{aligned}
  \end{equation}
 From this we find, for example,
 \begin{equation}\label{Diff-RHC-IHC}
   (ts)^{-1}\RHCr_{1,1}(t;x|s;y)- (xy)^{-1} \RHCl_{1,1}(t;x|s;y)=(q-q^{-1})\gun(s,t)\gun(y,x).
 \end{equation}
Thus, in contrast to the case of the left and the right Izergin determinants, the difference between
$\RHCl_{a,b}$ and $\RHCr_{a,b}$ can not be removed via simple multiplication of them by certain
sets of variables, like in \eqref{Diff-Mod-Izer}.

Below, to save space, we will combine the formulas for $\RHCl$ and $\RHCr$ into one. For instance,
the equations \eqref{RHC-IHC} and \eqref{2-RHC-IHC} can be written as follows:
 \begin{equation}\label{RHC-IHC-12}
  \RHClr_{a,b}(\bar t;\bar x|\bar s;\bar y)=(-q)^{\mp b}\sum
 \Izerrl_b(\bar s|\bar w_{\so}q^2)\Izerlr_a(\bar w_{\st}|\bar t)
  \Izerlr_b(\bar y|\bar w_{\so})\fun(\bar w_{\so},\bar w_{\st})\,.
 \end{equation}
The superscript $(l,r)$ on  ${\sf Z}_{a,b}$ means that the equation \eqref{RHC-IHC-12} is valid for $\RHCl_{a,b}$ and for $\RHCr_{a,b}$ separately. Choosing  the first or the second component of $(l,r)$ and the corresponding (up or down resp.) exponent of
$(-q)^{\mp b}$ in this equation, we obtain either \eqref{RHC-IHC} or \eqref{2-RHC-IHC}.

Similarly to the  case of the $GL(3)$-invariant $\RR$-matrix there exist slightly different representations, so-called  twin formula
for the highest coefficients:
 \begin{equation}\label{RHC-IHC-Tw}
  \RHClr_{a,b}(\bar t;\bar x|\bar s;\bar y)=(-q)^{\mp a}\sum
 \Izerrl_a(\bar w_{\st}|\bar xq^2)\Izerlr_a(\bar w_{\st}|\bar t)
  \Izerlr_b(\bar y|\bar w_{\so})\fun(\bar w_{\so},\bar w_{\st}).
 \end{equation}
All the notations are the same as in \eqref{RHC-IHC-12}.
This formula follows from the reduction properties of the Izergin determinants \eqref{K-red}.
Indeed, we have
\begin{multline}\label{Red-twin}
(-q)^{\pm b} \Izerlr_b(\bar s|\bar w_{\so}q^2)=(-q)^{\pm(a+b)}\Izerlr_{a+b}(\{\bar s,\bar x\}|\{\bar w_{\so}q^2, \bar xq^2\})\\
=(-q)^{\pm(a+b)}\Izerlr_{a+b}(\{\bar w_{\so},\bar w_{\st}\}|\{\bar w_{\so}q^2, \bar xq^2\})=(-q)^{\pm a}\Izerlr_a(\bar w_{\st}|\bar xq^2),
\end{multline}
where the superscript $(l,r)$ has the same meaning as in \eqref{RHC-IHC-12}. Due to \eqref{Red-twin} the
equivalence of \eqref{RHC-IHC-12} and \eqref{RHC-IHC-Tw} becomes evident.
Other representations for the highest coefficients in terms of sums over partitions are given
in section~\ref{S-DifRep}.

\section{Derivation of sum formulas for the highest coefficients\label{S-SerHC}}

In order to derive \eqref{RHC-IHC}, \eqref{2-RHC-IHC} we should calculate the  scalar product $S_{a,b}(\blac;\bmuc|\blab;\bmub)$
and  find rational coefficients of the products $\rrr_1(\blab)\rrr_3(\bmuc)$ and $\rrr_1(\blac)\rrr_3(\bmub)$.

\subsection{The coefficient of $\rrr_1(\blab)\rrr_3(\bmuc)$}
Here we calculate
the coefficient of $\rrr_1(\blab)\rrr_3(\bmuc)$. We start with the  dual Bethe vector in the
form \eqref{Or-form}.
For our goal it is enough to take only one term from the sum over partitions corresponding to $k=0$:
\begin{equation}\label{dBV-main-1}
\mathbb{C}^{a,b}(\blac;\bmuc) =\frac{\langle0|T_{32}(\bmuc)T_{21}(\blac)}{\lambda_2(\bmuc)\lambda_2(\blac)\fun(\bmuc,\blac)}
+\text{IT}\;,
\end{equation}
and we recall that IT means the terms that do not contribute to the coefficient we consider.
 Indeed, acting with $\mathbb{C}^{a,b}(\blac;\bmuc)$ on the  Bethe vector
we want to obtain the product $\rrr_3(\bmuc)$ over the
complete set $\bmuc$.  This is possible if and only if the product $T_{32}(\bmuc_{\st})$ in \eqref{Or-form} depends on the
complete set $\bmuc$, that is $\bmuc_{\st}=\bmuc$.  Hence,  $\blac_{\so}=\bmuc_{\so}
=\emptyset$, and all other terms in \eqref{Or-form} are not essential.

It remains to act successively with  $T_{21}(\blac)$ and $T_{32}(\bmuc)$ on the Bethe vector. Using \eqref{A-act21-2}
we obtain
 \begin{multline}\label{1-FA}
 \mathbb{C}^{a,b}(\blac;\bmuc)\mathbb{B}^{a,b}(\blab;\bmub)=\frac{(-q)^a}
 {\lambda_2(\bmuc)\fun(\bmuc,\blac)}\sum
\rrr_1(\bar\eta_{\so})\;
 \frac{\fun(\bar\eta_{\st},\bar\eta_{\so})\fun(\bar\xi_{\st},\bar\xi_{\so})}{\fun(\bar\xi_{\st},\bar\eta_{\so})} \num
 \times  \Izerr_a(q^{-2}\blac|\bar\eta_{\st})
  \Izerl_a(\bar\eta_{\so}|q^2\bar\xi_{\so})\Izerl_a(\bar\xi_{\so}|q^2\blac)
 \langle0|T_{32}(\bmuc)\mathbb{B}^{0,b}(\emptyset;\bar\xi_{\st})+\text{IT}\;.
 \end{multline}
Here $\bar\eta=\{\blac,\blab\}$ and $\bar\xi=\{\blac,\bmub\}$.  The sum is taken over partitions
$\bar\eta\Rightarrow\{\bar\eta_{\so},\bar\eta_{\st}\}$ and
$\bar\xi\Rightarrow\{\bar\xi_{\so},\bar\xi_{\st}\}$ with $\#\bar\eta_{\so}=\#\bar\xi_{\so}=a$.

The remaining action of
$T_{32}(\bmuc)$ on $\mathbb{B}^{0,b}(\emptyset;\bar\xi_{\st})$ should be calculated via
\eqref{A-act32-3}. This gives us
 \begin{multline}\label{1-SA}
 \mathbb{C}^{a,b}(\blac;\bmuc)\mathbb{B}^{a,b}(\blab;\bmub)=\frac{(-q)^a\rrr_3(\bmuc)}
 {\fun(\bmuc,\blac)}\sum
\rrr_1(\bar\eta_{\so})\;
 \frac{\fun(\bar\eta_{\st},\bar\eta_{\so})\fun(\bar\xi_{\st},\bar\xi_{\so})}{\fun(\bar\xi_{\st},\bar\eta_{\so})}
 \num
 \times  \Izerr_a(q^{-2}\blac|\bar\eta_{\st})
  \Izerl_a(\bar\eta_{\so}|q^2\bar\xi_{\so})\Izerl_a(\bar\xi_{\so}|q^2\blac)
  \Izerr_b(\bmuc|\bar\xi_{\st})+\text{IT}\;.
 \end{multline}

Now we should extract from the sum \eqref{1-SA} the terms proportional to $\rrr_1(\blab)$. For this we
should simply set $\bar\eta_{\so}=\blab$ and $\bar\eta_{\st}=\blac$. After elementary algebra
based on the use of \eqref{K-red} and \eqref{K-invers}
we arrive at
\beq{res-1}
 \mathbb{C}^{a,b}(\blac;\bmuc)\mathbb{B}^{a,b}(\blab;\bmub)=
 \frac{\rrr_1(\blab)\rrr_3(\bmuc)}
 {\fun(\bmuc,\blac)\fun(\bmub,\blab)}\;\RHCr_{a,b}(\blab;\blac|\bmub;\bmuc)+\text{IT}\;,
 \eeq
where $\RHCr_{a,b}$ has the following form:
\beq{Z2}
\RHCr_{a,b}(\blab;\blac|\bmub;\bmuc)=(-q)^a\sum \Izerl_a(\bar\xi_{\so}|q^2\blac)
\Izerr_a(\bar\xi_{\so}|\blab)\Izerr_b(\bmuc|\bar\xi_{\st})\fun(\bar\xi_{\st},\bar\xi_{\so}),
\eeq
where $\bar\xi=\{\blac,\bmub\}$,  and the sum is taken over partitions
$\bar\xi\Rightarrow\{\bar\xi_{\so},\bar\xi_{\st}\}$ with $\#\bar\xi_{\so}=a$. This coincides with \eqref{RHC-IHC-Tw} up to notations.

\subsection{The coefficient of $\rrr_1(\blac)\rrr_3(\bmub)$}

The calculation is very similar to the previous one. This time we start with the dual Bethe vector in the
form \eqref{Or-form2}.
Now we want to obtain the coefficient of $\rrr_1(\blac)\rrr_3(\bmub)$, therefore we have
\begin{equation}\label{dBV-main-2}
\mathbb{C}^{a,b}(\blac;\bmuc) =\frac{\langle0|T_{21}(\blac)T_{32}(\bmuc)}{\lambda_2(\bmuc)\lambda_2(\blac)\fun(\bmuc,\blac)}
+\text{IT}\;.
\end{equation}
We should act successively with   $T_{32}(\bmuc)$ and $T_{21}(\blac)$ on the  Bethe vector. Using \eqref{A-act32-2}
we obtain
 \begin{multline}\label{2-FA}
 \mathbb{C}^{a,b}(\blac;\bmuc)\mathbb{B}^{a,b}(\blab;\bmub)=
\frac{ (-q)^{-b}}{\lambda_2(\blac)\fun(\bmuc,\blac)}\sum \rrr_3(\bar\xi_{\so})\;
 \frac{\fun(\bar\xi_{\so},\bar\xi_{\st})\fun(\bar\eta_{\so},\bar\eta_{\st})}{\fun(\bar\xi_{\so},\bar\eta_{\st})} \num
 \times  \Izerr_b(q^{-2}\bmuc|\bar\eta_{\so})
 \Izerr_b(q^{-2}\bar\eta_{\so}|\bar\xi_{\so})\Izerl_b(\bar\xi_{\st}|q^2\bmuc)
 \langle0|T_{21}(\blac)\mathbb{B}^{a,0}(\bar\eta_{\st};\emptyset)+\text{IT}\;.
 \end{multline}
Here $\bar\eta=\{\bmuc,\blab\}$ and $\bar\xi=\{\bmuc,\bmub\}$.  The sum is taken over partitions
$\bar\eta\Rightarrow\{\bar\eta_{\so},\bar\eta_{\st}\}$ and
$\bar\xi\Rightarrow\{\bar\xi_{\so},\bar\xi_{\st}\}$ with $\#\bar\eta_{\so}=\#\bar\xi_{\so}=b$.

Now we act with $T_{21}(\blac)$ on $\mathbb{B}^{a,0}(\bar\eta_{\st};\emptyset)$ via \eqref{A-act21-3}
 \begin{multline}\label{2-SA}
 \mathbb{C}^{a,b}(\blac;\bmuc)\mathbb{B}^{a,b}(\blab;\bmub)=
\frac{ (-q)^{-b}\rrr_1(\blac)}{\fun(\bmuc,\blac)}\sum \rrr_3(\bar\xi_{\so})
 \frac{\fun(\bar\xi_{\so},\bar\xi_{\st})\fun(\bar\eta_{\so},\bar\eta_{\st})}{\fun(\bar\xi_{\so},\bar\eta_{\st})} \\
 \times  \Izerr_b(q^{-2}\bmuc|\bar\eta_{\so})
 \Izerr_b(q^{-2}\bar\eta_{\so}|\bar\xi_{\so})\Izerl_b(\bar\xi_{\st}|q^2\bmuc)
 \Izerl_a(\bar\eta_{\st}|\blac)+\text{IT}\;.
 \end{multline}
Setting here $\bar\xi_{\so}=\bmub$ and $\bar\xi_{\st}=\bmuc$ we obtain after trivial algebra
(and the use of \eqref{K-red}, \eqref{K-invers})
\beq{res-2}
 \mathbb{C}^{a,b}(\blac;\bmuc)\mathbb{B}^{a,b}(\blab;\bmub)=
 \frac{\rrr_1(\blac)\rrr_3(\bmub)}
 {\fun(\bmuc,\blac)\fun(\bmub,\blab)}\;\RHCl_{a,b}(\blac;\blab|\bmuc;\bmub)+\text{IT}\;,
 \eeq
where $\RHCl_{a,b}$ has the following form:
\beq{Z1}
\RHCl_{a,b}(\blac;\blab|\bmuc;\bmub)=(-q)^{-b}\sum \Izerr_b(q^{-2}\bmuc|\bar\eta_{\so})
\Izerl_b(\bmub|\bar\eta_{\so})\Izerl_a(\bar\eta_{\st}|\blac)\fun(\bar\eta_{\so},\bar\eta_{\st}),
\eeq
and $\bar\eta=\{\blab,\bmuc\}$. This coincides with \eqref{RHC-IHC} up to notations.

Thus, we have proved that the coefficient of the product
$\rrr_1(\blab)\rrr_3(\bmuc)$ is proportional to the function $\RHCr_{a,b}(\blab;\blac|\bmub;\bmuc)$, while
the coefficient of the product  $\rrr_1(\blac)\rrr_3(\bmub)$ is proportional to to the function
$\RHCl_{a,b}(\blac;\blab|\bmuc;\bmub)$.


\section{Properties and alternative expressions of $\RHCl$ and $\RHCr$ \label{sec:altern}}

In order to obtain a complete formula for the scalar product one should know some
properties of the highest coefficients. In particular, different representations
for $\RHClr_{a,b}$ are of great importance. The description of the residues of
$\RHClr_{a,b}$ in their poles, as well as some reduction properties, also are useful.
In this section we give a list of properties of the highest coefficients.

\subsection{Simple properties of the highest coefficients}

It is easy to see that both highest coefficients are symmetric with respect to all the permutations
of variables in  any of the four sets: $\bar t$, $\bar x$, $\bar s$, and $\bar y$.
It also follows immediately from the definitions
\eqref{RHC-IHC-12}, \eqref{RHC-IHC-Tw} that
\beq{Triv-pc}
\RHClr_{a,0}(\bar t;\bar x|\emptyset; \emptyset)=\Izerlr_a(\bar x|\bar t),\qquad
\RHClr_{0,b}(\emptyset; \emptyset|\bar s;\bar y)=\Izerlr_b(\bar y|\bar s).
\eeq
Using \eqref{K-scal} one can  easily see that the highest coefficients are invariant under
the rescaling of all arguments
\beq{Z-scal}
\RHClr_{a,b}(\alpha\bar t;\alpha\bar x|\alpha\bar s;\alpha\bar y)=\RHClr_{a,b}(\bar t;\bar x|\bar s;\bar y).
\eeq
In order to describe more sophisticated properties one should use different representations for
the highest coefficients.

\subsection{Different representations of $\RHCl$ and $\RHCr$\label{S-DifRep}}

Just like in the case of the $GL(3)$-invariant $\RR$-matrix there exist several representations for the highest coefficients
in terms of sums over partitions. The original formula \eqref{RHC-IHC-12} is
given in terms of the sums over partitions of the union of the sets $\{\bar x,\bar s\}=\bar w$. There are also
representations in terms of the sums over partitions of the unions of the sets $\{\bar tq^{-2},\bar y\}$,
$\{\bar t,\bar x\}$, and $\{\bar s,\bar y\}$. We give the complete list of these representations below.

\begin{itemize}
\item {\it Representations in terms of the partitions of $\{\bar tq^{-2},\bar y\}$. }
\end{itemize}
 \beq{Al-RHC-IHC}
 \RHClr_{a,b}(\bar t;\bar x|\bar s;\bar y)=(-q)^{\mp a}\fun(\bar y,\bar x)\fun(\bar s,\bar t)
   \sum \Izerrl_a(\bar tq^{-2}|\bar\eta_{\so}q^2)\Izerlr_a(\bar xq^{-2}|\bar\eta_{\so})\Izerlr_b(\bar\eta_{\st}|\bar s)\fun(\bar\eta_{\so},\bar\eta_{\st}).
      \eeq
Here $\bar\eta=\{\bar y,\;\bar tq^{-2}\}$. The sum is taken with respect to the partitions
$\bar\eta\Rightarrow\{\bar\eta_{\so},\bar\eta_{\st}\}$ with $\#\bar\eta_{\so}=a$ and $\#\bar\eta_{\st}=b$.

These representations also have a twin formula:
 \beq{Al-RHC-IHC-Tw}
 \RHClr_{a,b}(\bar t;\bar x|\bar s;\bar y)=(-q)^{\mp b}\fun(\bar y,\bar x)\fun(\bar s,\bar t)
   \sum \Izerrl_b(\bar\eta_{\st}|\bar yq^{2})\Izerlr_a(\bar xq^{-2}|\bar\eta_{\so})\Izerlr_b(\bar\eta_{\st}|\bar s)\fun(\bar\eta_{\so},\bar\eta_{\st}).
      \eeq
All the notations are the same as in \eqref{Al-RHC-IHC}. The twin-formula follows from \eqref{Al-RHC-IHC} due to the
identity $(-q)^{\mp a}\Izerrl_a(\bar tq^{-2}|\bar\eta_{\so}q^2)=
(-q)^{\mp b}\Izerrl_b(\bar\eta_{\st}|\bar yq^{2})$.

\begin{itemize}
\item {\it Representations in terms of the partitions of $\{\bar t,\bar x\}$. }
\end{itemize}
 \ba{GF}
 \RHClr_{a,b}(\bar t;\bar x|\bar s;\bar y)=\sum (-q)^{\pm n}\fun(\bar s,\bar t_{\so}) \fun(\bar y,\bar x_{\st})
 \fun(\bar t_{\so},\bar t_{\st})\fun(\bar x_{\st},\bar x_{\so})\\
 \times \Izerlr_n(\bar x_{\so}|\bar t_{\so})\Izerlr_{a-n}(\bar x_{\st}|\bar t_{\st}q^{-2})
 \Izerlr_{b+n}(\{\bar y,\bar t_{\so}q^{-2}\}|\{\bar s,\bar x_{\so}\}).
 \end{multline}
The sum is taken with respect to all partitions  $\bar t\Rightarrow\{\bar t_{\so},\bar t_{\st}\}$  and
$\bar x\Rightarrow\{\bar x_{\so},\bar x_{\st}\}$  with $\#\bar t_{\so}=
\#\bar x_{\so}=n$, $n=0,1,\dots,a$.

\begin{itemize}
\item {\it Representations in terms of the partitions of $\{\bar s,\bar y\}$. }
\end{itemize}
 \begin{multline}\label{S-GF}
  \RHClr_{a,b}(\bar t;\bar x|\bar s;\bar y)=\sum (-q)^{\pm n} \fun(\bar s_{\st},\bar t) \fun(\bar y_{\so},\bar x)
 \fun(\bar s_{\so},\bar s_{\st})\fun(\bar y_{\st},\bar y_{\so})\\
 \times \Izerlr_n(\bar y_{\so}|\bar s_{\so}) \Izerlr_{b-n}(\bar y_{\st}|\bar s_{\st}q^{-2})
  \Izerlr_{a+n}(\{\bar s_{\so},\bar x\}|\{\bar y_{\so}q^2,\bar t\}).
 \end{multline}
The sum is taken with respect to all partitions  $\bar s\Rightarrow\{\bar s_{\so},\bar s_{\st}\}$  and
$\bar y\Rightarrow\{\bar y_{\so},\bar y_{\st}\}$
 with $\#\bar s_{\so}=\#\bar y_{\so}=n$, $n=0,1,\dots,b$.

All the representations above follow from the original ones. In complete analogy with
the case of the $GL(3)$-invariant $\RR$-matrix the sums over partitions in \eqref{RHC-IHC-12} can be
presented as multiple contour integrals, where the integration contours surround the
points $\bar w=\{\bar s,\;\bar x\}$. Then moving these contours to the points
$\{\bar t, \bar x\}$ or $\{\bar s, \bar y\}$ (depending on the specific representation)
one obtains the equations \eqref{GF} or \eqref{S-GF}. We refer the reader to the work
\cite{BelPRS12a} for the details of this derivation.

Representations \eqref{GF} and \eqref{S-GF} allow us to prove a very important property
of $\RHClr$:
\beq{Z-invers}
 \RHClr_{b,a}(\bar s;\bar y|\bar tq^{-2};\bar xq^{-2})=\fun^{-1}(\bar y,\bar x)\fun^{-1}(\bar s,\bar t)
  \RHClr_{a,b}(\bar t;\bar x|\bar s;\bar y).
\eeq
This formula can be obtained by substitution of the $ \RHClr_{b,a}(\bar s;\bar y|\bar tq^{-2};\bar xq^{-2})$ into \eqref{GF}. This will give us \eqref{S-GF} for the   $\RHClr_{a,b}(\bar t;\bar x|\bar s;\bar y)$.

The property \eqref{Z-invers} immediately implies the representations \eqref{Al-RHC-IHC}, \eqref{Al-RHC-IHC-Tw}.
Indeed, one can easily check that using  \eqref{RHC-IHC-12} for
$\RHClr_{b,a}(\bar s;\bar y|\bar tq^{-2};\bar xq^{-2})$ we obtain the representations
 \eqref{Al-RHC-IHC} for $\RHClr_{a,b}(\bar t;\bar x|\bar s;\bar y)$.

One more property of the highest coefficients with respect to re-ordering of their arguments has the
following form:
\beq{Z-invers1}
\RHClr_{a,b;q^{-1}}(\bar t;\bar x|\bar s;\bar y)=\RHCrl_{b,a;q}(\bar y;\bar s|\bar x;\bar t)\,.
\eeq
Here we have added to the highest coefficients the subscripts $q$ and $q^{-1}$, in order to stress that
in the l.h.s. the function $\RHClr_{a,b}$ is evaluated with replacement
$q$ by $q^{-1}$. On the contrary, in the r.h.s. of \eqref{Z-invers1} the highest coefficient is evaluated at the same $q$,
but with replacement left by right, $a$ by $b$, and re-ordering of the arguments. Usually we omit the
additional subscript $q$ in the formulas. The proof of this property is given in appendix~\ref{A-Z-invers1}. Using \eqref{Z-invers1} and
\eqref{K-invers1} one can easily check that the representation \eqref{S-GF} follows from \eqref{GF}
after the replacement $q$ by $q^{-1}$.

We conclude this section by establishing  the behavior
of the highest coefficients as one of their arguments goes to infinity.
For this, it is convenient to use the representations \eqref{GF} and \eqref{S-GF}.  Due to \eqref{Klr-inf1}
and \eqref{Klr-inf2} one can easily convince himself that
 \begin{equation}\label{Zlr-inf1}
\begin{array}{lll}
 \RHCl_{a,b}(\bar t;\bar x|\bar s;\bar y)\to 0,& t_i\to\infty\quad\text{or}\quad s_j\to\infty,& \qquad i=1,\dots,a,\\
 \RHCl_{a,b}(\bar t;\bar x|\bar s;\bar y)\quad\text{is bounded},& x_i\to\infty\quad\text{or}\quad y_j\to\infty,&
 \qquad j=1,\dots,b.
 \end{array}
  \end{equation}
These equations together with the property \eqref{Z-invers1} yield
 \begin{equation}\label{Zlr-inf2}
 \begin{array}{lll}
 \RHCr_{a,b}(\bar t;\bar x|\bar s;\bar y)\to 0,& x_i\to\infty\quad\text{or} \quad y_j\to\infty,& \qquad i=1,\dots,a,\\
 \RHCr_{a,b}(\bar t;\bar x|\bar s;\bar y)\quad\text{is bounded},& t_i\to\infty\quad\text{or}\quad s_j\to\infty,& \qquad j=1,\dots,b.
 \end{array}
\end{equation}
%


\subsection{Special sums over partitions reducible to $\RHClr_{a,b}$}

All the sum formulas for $\RHClr_{a,b}$ involve the products of three Izergin determinants. There exits
more general formulas with three Izergin determinants, which are also reducible to the
highest coefficients. Such formulas are necessary for the derivation of sum representations for
the scalar product of the Bethe vectors. Below we give the list of these formulas.

Let $a\ge b$. Then
 \begin{equation}\label{Al-RHC-IHC-twin1}
  \sum
 \Izerrl_b(\bar t_{\so}|\bar y q^2)\Izerlr_b(\bar t_{\so}|\bar sq^2)
 \Izerlr_{a-b}(\bar\xi|\bar t_{\st})\fun(\bar t_{\st},\bar t_{\so})
   =(-q)^{\pm b}\frac{\RHClr_{a,b}(\bar t;\{\bar\xi,\bar y\}|\bar s;\bar yq^{-2})}{ \fun(\bar y,\bar t)\fun(\bar s,\bar t)},
        \end{equation}
%
 \begin{equation}
  \sum
 \Izerlr_b(\bar t_{\so}|\bar y q^2)\Izerrl_b(\bar t_{\so}|\bar sq^2)
 \Izerlr_{a-b}(\bar\xi|\bar t_{\st})\fun(\bar t_{\st},\bar t_{\so})
      =(-q)^{\pm b}\frac{\RHClr_{a,b}(\bar t;\{\bar\xi,\bar s\}|\bar y;\bar sq^{-2})}{ \fun(\bar y,\bar t)\fun(\bar s,\bar t)}.\label{Al-RHC-IHC-twin2}
     \end{equation}
Here the sum is taken over partitions $\bar t\Rightarrow\{\bar t_{\so},\bar t_{\st}\}$ with
$\#\bar t_{\so}=b$ and $\#\bar t_{\st}=a-b$.

Let now $a\le b$. Then
 \begin{equation}\label{Al-RHC-IHC-twin3}
  \sum
 \Izerrl_a(q^{-2}\bar t|\bar y_{\so} )\Izerlr_a(\bar xq^{-2}|\bar y_{\so})
 \Izerlr_{b-a}(\bar y_{\st}|\bar\xi)\fun(\bar y_{\so},\bar y_{\st})
   =(-q)^{\pm a}\;\frac{\RHClr_{a,b}(\bar tq^{2};\bar x|\{\bar\xi,\bar t\};\bar y)}{ \fun(\bar y,\bar t)\fun(\bar y,\bar x)},
     \end{equation}
 \begin{equation}
  \sum
 \Izerlr_a(q^{-2}\bar t|\bar y_{\so} )\Izerrl_a(\bar xq^{-2}|\bar y_{\so})
 \Izerlr_{b-a}(\bar y_{\st}|\bar\xi)\fun(\bar y_{\so},\bar y_{\st})
   =(-q)^{\pm a}\;\frac{\RHClr_{a,b}(\bar xq^{2};\bar t|\{\bar\xi,\bar x\};\bar y)}{ \fun(\bar y,\bar t)\fun(\bar y,\bar x)}. \label{Al-RHC-IHC-twin4}
     \end{equation}
Here the sum is taken over partitions $\bar y\Rightarrow\{\bar y_{\so},\bar y_{\st}\}$
with $\#\bar y_{\so}=a$ and  $\#\bar y_{\st}=b-a$.

All these formulas  follow from the  representations \eqref{Al-RHC-IHC}, \eqref{Al-RHC-IHC-Tw}
(see an example of the proof in appendix~\ref{A-sum3K}). Note also that due to \eqref{Z-invers1}
and \eqref{K-invers1} the equations \eqref{Al-RHC-IHC-twin3} and \eqref{Al-RHC-IHC-twin4} follow from respectively \eqref{Al-RHC-IHC-twin1} and \eqref{Al-RHC-IHC-twin2} via the replacement $q\to q^{-1}$.

\subsection{Poles of the highest coefficients}

The highest coefficients $\RHClr_{a,b}(\bar t;\bar x|\bar s;\bar y)$  have simple poles at $t_i=x_j$, $t_i=s_k$, $x_j=y_\ell$, and $s_k=y_\ell$. Similarly to the  case of the $GL(3)$-invariant $\RR$-matrix the corresponding residues
can be expressed in terms of $\RHClr_{a-1,b}(\bar t;\bar x|\bar s;\bar y)$ or $\RHClr_{a,b-1}(\bar t;\bar x|\bar s;\bar y)$.
In particular,
\beq{Rec-Z-triv1}
\Bigl.\RHClr_{a,b}(\bar t;\bar x|\bar s;\bar y)\Bigr|_{ s_b\to y_b}=
\fun(y_b,s_b)\fun( y_b,\bar s_{ b})\fun(\bar y_{ b}, y_b)\fun( y_b,\bar x)
\RHClr_{a,b-1}(\bar t;\bar x|\bar s_{ b};\bar y_{ b})+\text{reg},
\eeq
where $\text{reg}$ means regular part. We remind also that $\bar s_{ b}=\bar s\setminus s_{b}$ and
$\bar y_{ b}=\bar y\setminus  y_{b}$.

The residue at $t_a=x_a$ has similar form
\beq{Rec-Z-triv2}
\Bigl.\RHClr_{a,b}(\bar t;\bar x|\bar s;\bar y)\Bigr|_{ t_a\to x_a}=
 \fun(x_a,t_a)\fun( x_a,\bar t_a)\fun(\bar x_a, x_a)\fun(\bar s, x_a)
\RHClr_{a-1,b}(\bar t_a;\bar x_a|\bar s;\bar y)+\text{reg}.
\eeq
It is worth mentioning that equations \eqref{Rec-Z-triv1} and \eqref{Rec-Z-triv2} are not independent, because
they are related by the transforms \eqref{Z-invers} and  \eqref{Z-invers1}.

The formula for the residue at $s_b= t_a$ is slightly more sophisticated. Namely,
\begin{multline}\label{Rec-Z-nontriv}
\Bigl.\RHClr_{a,b}(\bar t;\bar x|\bar s;\bar y)\Bigr|_{ s_b\to t_a}=
 \fun(s_b, t_a) \fun( \bar s_{ b}, s_b)\fun( t_a, \bar t_{ a})\\
\times \sum_{p=1}^a \;\Izerlr_1(x_p| t_a)\fun(\bar x_{ p},x_p)
\RHClr_{a-1,b}(\bar t_{ a};\bar x_{ p}|\{\bar s_{ b}, x_p\};\bar y)+\text{reg}.
\end{multline}
Similarly the residue at $y_b= x_a$ is given by
\begin{multline}\label{Rec-Z-nontriv-d}
\Bigl.\RHClr_{a,b}(\bar t;\bar x|\bar s;\bar y)\Bigr|_{ y_b\to x_a}=
 \fun(y_b, x_a) \fun( \bar y_{ b}, y_b)\fun( x_a, \bar x_{ a})\\
\times \sum_{p=1}^b \Izerlr_1(x_a| s_p)\fun(s_{ p},\bar s_p)
\RHClr_{a,b-1}(\bar t;\{\bar x_{ a},s_p\}|\bar s_{ p};\bar y_b)+\text{reg}.
\end{multline}
The derivation of all these formulas is exactly the same as in the case of the $GL(3)$-invariant $\RR$-matrix, therefore we refer
the reader to \cite{BelPRS12a} for the corresponding proofs.
Note that the formulas \eqref{Rec-Z-nontriv} and  \eqref{Rec-Z-nontriv-d} are related by the properties
\eqref{Z-invers1}, \eqref{K-invers1}, and the transform $q\to q^{-1}$. 

\subsection{Multiple poles}

The residue formulas above imply multiple residue formulas. Namely, if $\#\bar z=n$, then it follows from
\eqref{Rec-Z-triv1} and \eqref{Rec-Z-triv2} that
\beq{red-2}
\lim_{\bar z'\to\bar z}\fun^{-1}(\bar z',\bar z)\RHClr_{a,b+n}(\bar t;\bar x|\{\bar s,\bar z\};\{\bar y,\bar z'\})
=\fun(\bar z,\bar x)\fun(\bar z,\bar s)\fun(\bar y,\bar z)\RHClr_{a,b}(\bar t;\bar x|\bar s;\bar y),
\eeq
and
\beq{red-1}
\lim_{\bar z'\to\bar z}\fun^{-1}(\bar z',\bar z)\RHClr_{a+n,b}(\{\bar t,\bar z\};\{\bar x,\bar z'\}|\bar s;\bar y)
=\fun(\bar z,\bar t)\fun(\bar x,\bar z)\fun(\bar s,\bar z)\RHClr_{a,b}(\bar t;\bar x|\bar s;\bar y).
\eeq
For $\#\bar z=1$ the equations \eqref{red-2}, \eqref{red-1} are direct corollaries of \eqref{Rec-Z-triv1},
\eqref{Rec-Z-triv2} respectively. Then one can use trivial induction over $n=\#\bar z$.

The residue formula \eqref{Rec-Z-nontriv}  implies the following reduction:
\begin{multline}\label{Rec-Z-nontriv-2}
\lim_{\bar z'\to\bar z} \fun^{-1}(\bar z,\bar z')\RHClr_{a,b}(\{\bar t,\bar z'\};\bar x|\{\bar s,\bar z\};\bar y)=
\fun( \bar s, \bar z)\fun( \bar z, \bar t)\\
\times \sum \Izerlr_n(\bar x_{\so}| \bar z)\fun(\bar x_{\st},\bar x_{\so})
\RHClr_{a-n,b}(\bar t;\bar x_{ \st}|\{\bar s, \bar x_{\so}\};\bar y).
\end{multline}
The sum is taken with respect to the  partitions $\bar x\Rightarrow\{\bar x_{\so},\bar x_{\st}\}$ with
$\#\bar x_{\so}=n$.

Similarly, starting from \eqref{Rec-Z-nontriv-d}  one can find that
\begin{multline}\label{Rec-Z-nontriv-22}
\lim_{\bar z'\to\bar z} \fun^{-1}(\bar z,\bar z')\RHClr_{a,b}(\bar t;\{\bar x,\bar z'\}|\bar s;\{\bar y,\bar z\})=
\fun( \bar y, \bar z)\fun( \bar z, \bar x)\\
\times \sum \Izerlr_n(\bar z|\bar s_{\so})\fun(\bar s_{\so},\bar s_{\st})
\RHClr_{a,b-n}(\bar t;\{\bar  x, \bar s_{ \so}\}|\bar s_{ \st};\bar y).
\end{multline}
Here the sum is taken with respect to the partitions $\bar s\Rightarrow\{\bar s_{\so},\bar s_{\st}\}$ with
$\#\bar s_{\so}=n$.

For $\#\bar z=1$ the equations \eqref{Rec-Z-nontriv-2}, \eqref{Rec-Z-nontriv-22} follow immediately from \eqref{Rec-Z-nontriv}, \eqref{Rec-Z-nontriv-d} respectively. Then one can proceed via induction over $n=\#\bar z$, using
the identities \eqref{Sym-Part-old1}, \eqref{Sym-Part-old2} (see details in appendix~\ref{A-Pr-red}).

\subsection{Reductions}

The reduction formulas \eqref{Rec-Z-nontriv-2}, \eqref{Rec-Z-nontriv-22} can be transformed. Namely, one can
apply the transform \eqref{Z-invers} to these equations (see
appendix~\ref{A-Doc-red}). In this way  we arrive at the following reductions
\begin{equation}\label{dec-2}
\RHClr_{a,b}(\{\bar t,q^2\bar z\};\bar x|\{\bar s,\bar z\};\bar y)=\sum \Izerlr_n(\bar y_{\so}|\bar z)
\RHClr_{a,b-n}(\{\bar t,q^2\bar y_{\so}\};\bar x|\bar s;\bar y_{\st})
\fun(\bar y_{\st},\bar y_{\so})\fun(\bar y_{\so},\bar x)\fun(\bar y_{\so},\bar s).
\end{equation}
The sum is taken with respect to the partitions $\bar y\Rightarrow\{\bar y_{\so},\bar y_{\st}\}$ with
$\#\bar y_{\so}=n$. One more reduction has the form
\begin{equation}\label{dec-1}
\RHClr_{a,b}(\bar t;\{\bar x,\bar z\}|\bar s;\{\bar y,\bar z q^{-2}\})=\sum
\Izerlr_n(\bar z|\bar t_{\so})
\RHClr_{a-n,b}(\bar t_{\st};\bar x|\bar s;\{\bar y,\bar t_{\so}q^{-2}\})
\fun(\bar t_{\so},\bar t_{\st})\fun(\bar x,\bar t_{\so})\fun(\bar s,\bar t_{\so}).
\end{equation}
Here the sum is taken with respect to the partitions $\bar t\Rightarrow\{\bar t_{\so},\bar t_{\st}\}$ with
$\#\bar t_{\so}=n$.

These formulas have special cases, when the highest coefficients degenerate into the products
of the Izergin determinants. In particular, if $b\le a$ and $n=b$, then in \eqref{dec-2}
$\bar s=\emptyset$ and $\bar y_{\st}=\emptyset$. The equation \eqref{dec-2} turns into
\begin{equation}\label{dec-2-pc}
\RHClr_{a,b}(\{\bar t,q^2\bar z\};\bar x|\bar z;\bar y)=\fun(\bar y,\bar x)\Izerlr_b(\bar y|\bar z)
\Izerlr_{a}(\bar x|\{\bar t,q^2\bar y\}).
\end{equation}
Similarly, if $a\le b$ and $n=a$, then  in \eqref{dec-1} $\bar x=\emptyset$ and $\bar t_{\st}=\emptyset$. The equation \eqref{dec-1} turns into
\begin{equation}\label{dec-1-pc}
\RHClr_{a,b}(\bar t;\bar z|\bar s;\{\bar y,\bar z q^{-2}\})=\fun(\bar s,\bar t)\Izerlr_a(\bar z|\bar t)
\Izerlr_{b}(\{\bar y,\bar tq^{-2}\}|\bar s).
\end{equation}
We draw the attention of the reader  that \eqref{dec-1-pc} is the image of \eqref{dec-2-pc} under the replacement
of $q$ by $q^{-1}$ and the use of eqs. \eqref{Z-invers1}
and \eqref{K-invers1}.

\subsection{Summation identity for the highest coefficients\label{S-SIHC}}

The equations \eqref{dec-2}, \eqref{dec-1} can be considered as summation identities, which allow one to
express certain sums  involving $\Izerlr$ and $\RHClr$ in terms of the highest coefficient $\RHClr$. In these
identities one takes a sum with respect to partitions of one set of variables. There exist more sophisticated
identities of similar type, where one takes a sum with respect to partitions of two sets of variables. In this
section we give one of such the identities. It plays very important role in the calculation of scalar products.

\begin{prop}\label{MainProp1}
Let $a$, $b$, $n$, $p$ be non-negative integers and $p\le b$. Let $\bar t$, $\bar x$, $\bar s$, $\bar y$, $\bar w$, $\bar z$ be
six sets of generic complex variables with cardinalities
\begin{equation}\label{cardi-sets}
\begin{array}{lll}
\#\bar t=a, &\#\bar x=a,& \#\bar z=n,\\
\#\bar s=b, &\#\bar y=p, &\#\bar w=b-p.
\end{array}
\end{equation}
Then
\begin{multline}\label{Sum-F}
\fun(\bar \xi,\bar y)\RHClr_{a,b}(\bar t;\bar x|\bar s;\{\bar y,\bar w\})=\sum (-q)^{\mp k}
\Izerrl_p(\{\bar s_{\so}q^{-2},\bar\xi_{\so}\}|\bar y)
\RHClr_{a,b-k}(\bar t;\bar x|\bar s_{\st};\{\bar w,\bar\xi_{\so}\}) \\
\times \fun(\bar s_{\so},\bar s_{\st})\fun(\bar \xi_{\st},\bar \xi_{\so})
\fun(\bar y,\bar s_{\so})\fun(\bar w,\bar s_{\so})\fun^{-1}(\bar s_{\so},\bar z).
\end{multline}
Here $\bar\xi$ is a union of two sets: $\bar\xi=\{\bar xq^{-2},\bar zq^{-2}\}$. The sum is taken
over partitions of the set $\bar s\Rightarrow\{\bar s_{\so},\bar s_{\st}\}$ with $\#\bar s_{\so}=k\in[0,\dots,p]$
and the set $\bar \xi\Rightarrow\{\bar\xi_{\so},\bar\xi_{\st}\}$ with $\#\bar\xi_{\so}=p-k$.
\end{prop}

The proof of this identity is given in appendix~\ref{A-PMP}.

\section*{Conclusion}

In this paper we have obtained several explicit representations for the highest coefficients $\RHClr$  for integrable models based on $GL(3)$ trigonometric $R$-matrix and found
their properties. Of course, this result is only a first step towards the calculation of the scalar products of Bethe vectors and then of the correlation functions for local operators. The calculation of the scalar products will be done in
our forthcoming publication, where we are going to use the present results. Indeed,
as we have explained in section~\ref{sec:sumHC},
the scalar product can be presented as a sum with respect to partitions of the Bethe parameters \eqref{scal}. The rational
coefficients $W_{\text{part}}$ in this equation are proportional to the product of the left and the right highest
coefficients. Hence the knowledge of highest coefficients is essential in the calculation of the scalar product.
To stress this fact, we announce  an explicit expression for $W_{\text{part}}$, that will be proved in our forthcoming publication.

\begin{prop}
The scalar product of two Bethe vectors \eqref{Def-SP} is given by equation \eqref{scal}. For
a fixed partition with $\#\blac_{\so}=\#\blab_{\so}=k$ and $\#\bmuc_{\so}=\#\bmub_{\so}=n$, (where $k=0,\dots,a$ and
$n=0,\dots,b$), the rational coefficient $W_{\text{part}}$ has the form
 \begin{multline}\label{W-Reshet}
W_{\text{part}}\begin{pmatrix}\blac_{\st},\blab_{\st},\blac_{\so},\blab_{\so}\\
\bmuc_{\so},\bmub_{\so},\bmuc_{\st},\bmub_{\st}\end{pmatrix}=\fun(\blab_{\st},\blab_{\so}) \fun(\blac_{\so},\blac_{\st})\fun(\bmub_{\so},\bmub_{\st})\fun(\bmuc_{\st},\bmuc_{\so})
\fun(\bmuc_{\so},\blac_{\so}) \fun(\bmub_{\st},\blab_{\st})\num
\times
\RHCl_{a-k,n}(\blac_{\st};\blab_{\st}|\bmuc_{\so};\bmub_{\so}) \;\RHCr_{k,b-n}(\blab_{\so};\blac_{\so}|\bmub_{\st};\bmuc_{\st})\;.
 \end{multline}
\end{prop}

In the scaling limit $u=e^{\varepsilon u'}$, $v=e^{\varepsilon v'}$, $q=e^{\varepsilon c/2}$,  $\varepsilon\to0$, the trigonometric $\RR$-matrix goes to the $GL(3)$-invariant $\RR$-matrix.  Then the functions
$\RHCl$ and $\RHCr$ coincide, and this formula turns into the representation obtained in \cite{Res86}. The last one
was already found to be useful for the analysis of form factors of local operators in $GL(3)$-invariant integrable
models. We hope that the explicit representation \eqref{W-Reshet} will be also fruitful for the study of integrable
models based on the $q$-deformed $GL(3)$ symmetry.

\section*{Acknowledgements}
We warmly thank S. Belliard for his contribution at the early stage of this work.
Work of S.P. was supported in part by RFBR grant 11-01-00962-a and  grant
of Scientific Foundation of NRU HSE 12-09-0064. E.R. was supported by ANR Project
DIADEMS (Programme Blanc ANR SIMI1 2010-BLAN-0120-02).
N.A.S. was  supported by the Program of RAS Basic Problems of the Nonlinear Dynamics,
grants RFBR-11-01-00440-a, RFBR-13-01-12405-ofi-m2, SS-2484.2014.1.

\appendix

\section{Properties of Izergin determinants\label{A-PID}}

Most of the properties of the left and right Izergin determinants easily follow
directly from their definitions \eqref{Izer}, \eqref{Mod-Izer}. We give below a list of these properties.
We remind that the  superscript $(l,r)$ on $\Izer$ means that the equality is valid for $\Izerl$ and for $\Izerr$ with appropriate choice of component (first/up or second/down) throughout the equality.

Initial condition:
 \begin{equation}\label{K-init}
\Izerl_{1}(\bar x|\bar y) = x\;\gun(x,y),\qquad \Izerr_{1}(\bar x|\bar y) = y\;\gun(x,y).
\end{equation}
Scaling:
 \begin{equation}\label{K-scal}
\Izerlr_{n}(\alpha\bar x|\alpha\bar y) = \Izerlr_{n}(\bar x|\bar y).
\end{equation}
Reduction:
 \begin{equation}\label{K-red}
\Izerlr_{n+1}(\{\bar x, q^{-2}z\}|\{\bar y,z\}) =\Izerlr_{n+1}(\{\bar x, z\}|\{\bar y,q^{2}z\})=-q^{\mp 1}\Izerlr_{n}(\bar x|\bar y).
\end{equation}
Inverse order of arguments:
 \begin{align}\label{K-invers}
  \Izerlr_{n}( q^{-2}\bar x|\bar y)
&= (-q)^{\mp n} \fun^{-1}(\bar y,\bar x) \Izerrl_{n}(\bar y|\bar x)\,,\\
\Izerlr_{n;q^{-1}}(\bar x|\bar y)&  =\Izerrl_{n;q}(\bar y|\bar x)\,,\label{K-invers1}
\end{align}
where $\Izerlr_{n;q^{-1}}$ means $\Izerlr_{n}$ with $q$ replaced by $q^{-1}$.
As for $\RHClr_{a,b}$ and relation \eqref{Z-invers1}, we have put an additional index $q^{-1}$ or $q$ in \eqref{K-invers1} to stress this replacement.

Residues in the poles:
 \begin{equation}\label{K-Res}
\Bigl.\Izerlr_{n+1}(\{\bar x,z\}|\{\bar y,z'\})\Bigr|_{z'\to z}= \fun(z,z')
\fun(z,\bar y)\fun(\bar x,z) \Izerlr_{n}(\bar x|\bar y)+{\rm reg},
\end{equation}
where ${\rm reg}$ means the regular part.

Behavior at infinity:
 \begin{equation}\label{Klr-inf1}
\begin{array}{ll}
 \Izerl_{n}(\bar x|\bar y)\sim y_i^{-1},& y_i\to\infty,\\
 \Izerr_{n}(\bar x|\bar y)\sim x_i^{-1},& x_i\to\infty,
 \end{array}\qquad
  i=1,\dots,n,
\end{equation}
and
 \begin{equation}\label{Klr-inf2}
 \begin{array}{ll}
 \Izerl_{n}(\bar x|\bar y)\quad\text{is bounded},& x_i\to\infty,\\
 \Izerr_{n}(\bar x|\bar y)\quad\text{is bounded},& y_i\to\infty,
 \end{array}
 \qquad i=1,\dots,n.
\end{equation}

\begin{prop}
Let $\#\bar x=\#\bar y=n$ and $\#\bar z=\#\bar z'=m$. Then
\beq{Mult-poleK}
\lim_{\bar z'\to\bar z}\fun^{-1}(\bar z,\bar z')\Izerlr_{n+m}(\{\bar x,\bar z\}|\{\bar y,\bar z'\})=
\fun(\bar x,\bar z)\fun(\bar z,\bar y)\Izerlr_{n}(\bar x|\bar y).
\end{equation}
\end{prop}

{\sl Proof}. Using \eqref{K-invers} we have
\beq{MPK-1}
\Izerlr_{n+m}(\{\bar x,\bar z\}|\{\bar y,\bar z'\})=(-q)^{\mp(m+n)}
\Izerrl_{n+m}(\{\bar y,\bar z'\}|\{\bar xq^2,\bar zq^2\})\fun(\bar x,\bar y)
\fun(\bar x,\bar z')\fun(\bar z,\bar y)\fun(\bar z,\bar z').
\end{equation}
The limit $\bar z'\to\bar z$ becomes trivial, and using successively \eqref{K-red}, \eqref{K-invers} we arrive at
\eqref{Mult-poleK}.

The Izergin determinants satisfy also summation identities.

\begin{lemma}\label{main-ident}
Let $\bar\gamma$, $\bar\alpha$ and $\bar\beta$ be three sets of complex variables with $\#\alpha=m_1$,
$\#\beta=m_2$, and $\#\gamma=m_1+m_2$. Then
\begin{equation}\label{Sym-Part-old1}
  \sum
 \Izerlr_{m_1}(\bar\gamma_{\so}|\bar \alpha)\Izerrl_{m_2}(\bar \beta|\bar\gamma_{\st})\fun(\bar\gamma_{\st},\bar\gamma_{\so})
 = (-q)^{\mp m_1}\fun(\bar\gamma,\bar \alpha) \Izerrl_{m_1+m_2}(\{\bar \alpha q^{-2},\bar \beta\}|\bar\gamma).
 \end{equation}
The sum is taken with respect to all partitions of the set $\bar\gamma\Rightarrow\{\bar\gamma_{\so},\bar\gamma_{\st}\}$ with $\#\bar\gamma_{\so}=m_1$ and $\#\bar\gamma_{\st}=m_2$.
Due to \eqref{K-invers} the equation \eqref{Sym-Part-old1} can be also written in the form
\begin{equation}\label{Sym-Part-old2}
  \sum
 \Izerlr_{m_1}(\bar\gamma_{\so}|\bar \alpha)\Izerrl_{m_2}(\bar \beta|\bar\gamma_{\st})\fun(\bar\gamma_{\st},\bar\gamma_{\so})
 = (-q)^{\pm m_2}\fun(\bar \beta,\bar\gamma) \Izerlr_{m_1+m_2}(\bar\gamma|\{\bar \alpha,\bar \beta q^2\}).
 \end{equation}
\end{lemma}

This statement is a simple corollary of Lemma~1 of the work \cite{BelPRS12b}.

\section{Some proofs}

\subsection{Proof of \eqref{Z-invers1}\label{A-Z-invers1}}

We take the representation \eqref{RHC-IHC-12} and replace there $q$ by $q^{-1}$. Using \eqref{K-invers1}
we obtain
 \begin{equation}\label{A-RHC-IHC-12}
  \RHClr_{a,b;q^{-1}}(\bar t;\bar x|\bar s;\bar y)=(-q)^{\mp b}\sum
 \Izerlr_b(\bar w_{\so}q^{-2}|\bar s)\Izerrl_a(\bar t|\bar w_{\st})
  \Izerrl_b(\bar w_{\so}|\bar y)\fun(\bar w_{\st},\bar w_{\so})\,,
 \end{equation}
where $\bar w=\{\bar x,\bar s\}$. Here we have used the evident property of the function $\fun(x,y)$ under the replacement
$q\to q^{-1}$: $\fun_{q^{-1}}(x,y)=\fun(y,x)$. Replacing $\bar w_{\so}\leftrightarrow\bar w_{\st}$ we find that
the equation \eqref{A-RHC-IHC-12} coincides with the representation \eqref{RHC-IHC-Tw} for $\RHCrl_{b,a}(\bar y;\bar s|\bar x;\bar t)$.
\qed 

\subsection{Proof of \eqref{Al-RHC-IHC-twin1}\label{A-sum3K}}

Consider the highest coefficient $\RHClr_{a,b}(\bar t;\{\bar\xi,\bar y'\}|\bar s;\bar yq^{-2})$  for $a\ge b$. Using
\eqref{Al-RHC-IHC-Tw} we obtain
 \begin{multline}\label{Pr-3K}
(-q)^{\pm b}\frac{\RHClr_{a,b}(\bar t;\{\bar\xi,\bar y'\}|\bar s;\bar yq^{-2})}{ \fun(\bar y,\bar t)\fun(\bar s,\bar t)}
= \frac{\fun(\bar yq^{-2},\bar\xi) \fun(\bar yq^{-2},\bar y')}{\fun(\bar y,\bar t)}\\
\times\sum
 \Izerrl_b(\bar\eta_{\st}|\bar y)\Izerlr_b(\bar\eta_{\st}|\bar s)\Izerlr_{a}(\{\bar\xi,\bar y'\}|\bar\eta_{\so}q^2)
 \fun(\bar\eta_{\so},\bar\eta_{\st})\,.
        \end{multline}
Here $\bar\eta=\{\bar tq^{-2},\bar yq^{-2}\}$,  $\#\bar t=a$, $\#\bar s=\#\bar y=\#\bar y'=b$,
$\#\bar\xi=a-b$, and $\#\bar\eta_{\so}=a$.  Consider the limit $\bar y'\to\bar y$. Then the product
$\fun(\bar yq^{-2},\bar y')=\fun^{-1}(\bar y',\bar y)$ vanishes. However the Izergin determinant
$\Izerlr_{a}(\{\bar\xi,\bar y'\}|\bar\eta_{\so}q^2)$ may have poles at $y'_i=y_i$. Evidently, the
complete compensation of the vanishing product $\fun^{-1}(\bar y',\bar y)$ occurs if and only
if $\bar yq^{-2}\subset\bar\eta_{\so}$. Then we can set $\bar\eta_{\so}=\{\bar yq^{-2},\bar t_{\st}q^{-2}\}$
and $\bar\eta_{\st}=\bar t_{\so}q^{-2}$. Substituting this into \eqref{Pr-3K} we obtain
 \begin{multline}\label{Pr-3K-1}
(-q)^{\pm b}\frac{\RHClr_{a,b}(\bar t;\{\bar\xi,\bar y\}|\bar s;\bar yq^{-2})}{ \fun(\bar y,\bar t)\fun(\bar s,\bar t)}
= \lim_{\bar y'\to\bar y}\fun^{-1}(\bar y',\bar y)\fun^{-1}(\bar\xi,\bar y) \fun^{-1}(\bar y,\bar t)\\
\times\sum
 \Izerrl_b(\bar t_{\so}q^{-2}|\bar y)\Izerlr_b(\bar t_{\so}q^{-2}|\bar s)
 \Izerlr_{a}(\{\bar\xi,\bar y'\}|\{\bar t_{\st},\bar y\})
 \fun(\bar y,\bar t_{\so}) \fun(\bar t_{\st},\bar t_{\so})\;,
        \end{multline}
where the sum is taken over partitions $\bar t\Rightarrow\{\bar t_{\so},\bar t_{\st}\}$
with $\#\bar t_{\so}=b$.
It remains to take the limit via \eqref{Mult-poleK}, and we arrive at \eqref{Al-RHC-IHC-twin1}.\qed

\subsection{Proof of \eqref{Rec-Z-nontriv-2}\label{A-Pr-red}}

Let \eqref{Rec-Z-nontriv-2} be valid for $\#\bar z =n-1$. Consider the case $\#\bar z =n$.
Taking the limit successively first for $\bar z'_n\to \bar z_n$ and then for
$z'_n\to z_n$ we obtain
\begin{multline}\label{Pr-rec1}
\lim_{\bar z'\to\bar z} \fun^{-1}(\bar z,\bar z')\RHClr_{a,b}(\{\bar t,\bar z'\};\bar x|\{\bar s,\bar z\};\bar y)=
\fun( \bar s, \bar z_n)\fun( \bar z_n, \bar t)
\sum \Izerlr_{n-1}(\bar x_{\so}| \bar z_n)\fun(\bar x_{\st},\bar x_{\so})\num
\times
\lim_{\bar z_n'\to\bar z_n} \fun^{-1}(z_n,z'_n)
\RHClr_{a-n+1,b}(\{\bar t,z'_n\};\bar x_{ \st}|\{\bar s, \bar x_{\so},z_n\};\bar y)\\
=\fun( \bar s, \bar z)\fun( \bar z, \bar t)\sum
\RHClr_{a-n,b}(\bar t;\bar x_{ \rm ii}|\{\bar s, \bar x_{\so},\bar x_{\rm i}\};\bar y)\\
\times\Izerlr_{n-1}(\bar x_{\so}| \bar z_n)\fun(\bar x_{\st},\bar x_{\so})
\Izerlr_{1}(\bar x_{\rm i}| z_n)\fun(\bar x_{\rm ii},\bar x_{\rm i})\fun(\bar x_{\so},  z_n).
\end{multline}
Here we first divide $\bar x$ into subsets $\{\bar x_{\so},\bar x_{\st}\}$ with $\#\bar x_{\so}=n-1$, and then
split the subset $\bar x_{\st}$ into sub-subsets $\{\bar x_{\rm i},\bar x_{\rm ii}\}$ with $\#\bar x_{\rm i}=1$.
Setting $\{\bar x_{\rm i},\bar x_{\so}\}=\bar x_{0}$ and using
$\Izerlr_{1}(\bar x_{\rm i}| z_n)=(-q)^{\mp 1}\fun(\bar x_{\rm i}, z_n)\Izerrl_{1}( z_nq^{-2}|\bar x_{\rm i})$
we find
\begin{multline}\label{Pr-rec2}
\lim_{\bar z'\to\bar z} \fun^{-1}(\bar z,\bar z')\RHClr_{a,b}(\{\bar t,\bar z'\};\bar x|\{\bar s,\bar z\};\bar y)=
\fun( \bar s, \bar z)\fun( \bar z, \bar t)\num
\times\sum
\RHClr_{a-n,b}(\bar t;\bar x_{ \rm ii}|\{\bar s, \bar x_{0}\};\bar y)\fun(\bar x_{\rm ii},\bar x_{\rm 0})
\fun(\bar x_{0},  z_n)\\
\times (-q)^{\mp 1}\Izerlr_{n-1}(\bar x_{\so}| \bar z_n)\Izerrl_{1}( z_nq^{-2}|\bar x_{\rm i})\fun(\bar x_{\rm i},\bar x_{\so}).
\end{multline}
Applying  lemma~\ref{main-ident} to the last line of \eqref{Pr-rec2} we can take the sum with respect to the
partitions $\bar x_0\Rightarrow\{\bar x_{\rm i},\bar x_{\so}\}$, that finally gives
\begin{multline}\label{Pr-rec3}
\lim_{\bar z'\to\bar z} \fun^{-1}(\bar z,\bar z')\RHClr_{a,b}(\{\bar t,\bar z'\};\bar x|\{\bar s,\bar z\};\bar y)=
\fun( \bar s, \bar z)\fun( \bar z, \bar t)\num
\times\sum
\Izerlr_{n}(\bar x_{0}| \bar z)\fun(\bar x_{\rm ii},\bar x_{\rm 0})\RHClr_{a-n,b}(\bar t;\bar x_{ \rm ii}|\{\bar s, \bar x_{0}\};\bar y).
\end{multline}
\qed
\subsection{Proof of \eqref{dec-1}\label{A-Doc-red}}

Let us simply write \eqref{Rec-Z-nontriv-22} replacing $a$ by $b$ and setting:
$\bar t=q^2\bar u$, $\bar x=q^2\bar v$, $\bar s=\bar\alpha$,
and $\bar y=\bar\beta$
\begin{multline}\label{Doc-Rec1}
\lim_{\bar z'\to\bar z} \fun^{-1}(\bar z,\bar z')\RHClr_{b,a}(q^2\bar u;\{q^2\bar v,\bar z'\}|\bar\alpha;\{\bar\beta,\bar z\})=
\fun( \bar\beta, \bar z)\fun( \bar z, q^2\bar v)\\
\times \sum \Izerlr_n(\bar z|\bar\alpha_{\so})\fun(\bar\alpha_{\so},\bar\alpha_{\st})
\RHClr_{b,a-n}(q^2\bar u;\{q^2\bar v, \bar\alpha_{ \so}\}|\bar\alpha_{ \st};\bar\beta)\;,
\end{multline}
 where the sum is taken over partitions $\bar\alpha\Rightarrow\{\bar\alpha_{\so},\bar\alpha_{\st}\}$
with $\#\bar\alpha_{\so}=n$.
Now we can use \eqref{Z-scal} and \eqref{Z-invers} to transform $\RHClr_{b,a}$ and $\RHClr_{b,a-n}$ in \eqref{Doc-Rec1}.
We have
\begin{equation}\label{Doc-Rec2}
\lim_{\bar z'\to\bar z} \fun^{-1}(\bar z,\bar z')\RHClr_{b,a}(q^2\bar u;\{q^2\bar v,\bar z'\}|\bar\alpha;\{\bar\beta,\bar z\})
=\frac{\fun(\bar\beta,\bar z)\;\RHClr_{a,b}(\bar\alpha;\{\bar\beta,\bar z\}|\bar u;\{\bar v,q^{-2}\bar z\})}
{\fun(\bar u,\bar\alpha)\fun(\bar v,\bar z)\fun(\bar v,\bar\beta)},
\end{equation}
and
\begin{equation}\label{Doc-Rec3}
\RHClr_{b,a-n}(q^2\bar u;\{q^2\bar v, \bar\alpha_{ \so}\}|\bar\alpha_{ \st};\bar\beta)
=\frac{\fun(\bar\beta,\bar\alpha_{ \so}) \;\RHClr_{a-n,b}(\bar\alpha_{ \st};\bar\beta|\bar u;\{\bar v, q^{-2}\bar\alpha_{ \so}\})}
{\fun(\bar v,\bar\beta)\fun(\bar u,\bar\alpha_{\st})}.
\end{equation}
Substituting all this into \eqref{Doc-Rec1} after evident cancelations we obtain
\begin{multline}\label{Doc-Rec4}
\RHClr_{a,b}(\bar\alpha;\{\bar\beta,\bar z\}|\bar u;\{\bar v,q^{-2}\bar z\})=
\sum \Izerlr_n(\bar z|\bar\alpha_{\so})\fun(\bar\alpha_{\so},\bar\alpha_{\st})\\
\times
\fun(\bar\beta,\bar\alpha_{ \so})\fun(\bar u,\bar\alpha_{\so})
\RHClr_{a-n,b}(\bar\alpha_{ \st};\bar\beta|\bar u;\{\bar v, q^{-2}\bar\alpha_{ \so}\})\;.
\end{multline}
It remains to set $\bar\alpha=\bar t$, $\bar\beta=\bar x$, $\bar u=\bar s$, $\bar v=\bar y$, and we arrive at
\eqref{dec-1}. \qed

\subsection{Proof of Proposition~\ref{MainProp1}\label{A-PMP}}

We use induction over $p$.
Denote the l.h.s. and the r.h.s. of \eqref{Sum-F} by $F^{(l,r)}_{a,b,p,n}
(\bar t;\bar x|\bar s;\bar w;\bar y|\bar z)$ and $\tilde F^{(l,r)}_{a,b,p,n}
(\bar t;\bar x|\bar s;\bar w;\bar y|\bar z)$ respectively.
For $p=0$ the equation \eqref{Sum-F} is trivial. Indeed, since $k\le p$ we obtain
that $\bar s_{\so}=\bar\xi_{\so}=\bar y =\emptyset$ for $p=0$. Hence, the sum over
partitions in the r.h.s. of \eqref{Sum-F} reduces to the one term, and both sides of this equation give
$\RHClr_{a,b}(\bar t;\bar x|\bar s;\bar w)$.

Now let \eqref{Sum-F} be valid for $\#\bar y=p-1$ and arbitrary $a$, $b$, and $n$:
\begin{equation}\label{Ind-ass}
F^{(l,r)}_{a,b,p-1,n}(\bar t;\bar x|\bar s;\bar w;\bar y|\bar z)=
\tilde F^{(l,r)}_{a,b,p-1,n}(\bar t;\bar x|\bar s;\bar w;\bar y|\bar z),\qquad \forall a,b,n\,.
\end{equation}
The general strategy of the proof is the following. We consider both sides of this equation
at $\#\bar y=p$ as
functions of $y_p$, the other variables being fixed. Obviously $F$ and $\tilde F$ are rational functions of $y_p$. We first establish that these functions have their poles in the same points and then  prove
that due to the induction assumption \eqref{Ind-ass} the residues in these poles coincide. Then it means that the difference
$F-\tilde F$ is a polynomial in $y_p$. Finally taking into account the
behavior of this polynomial at $y_p\to\infty$ and $y_p=0$ we conclude that it is identically equal to zero.

Obviously, the function
$F^{(l,r)}_{a,b,p,n}$ has poles at $y_p=\xi_\ell$, $\ell=1,\dots,a+n$ due to the factor $\fun(\bar\xi,\bar y)$. The highest coefficient $\RHClr_{a,b}$
has additional poles at $y_p=s_i$, $i=1,\dots,b$. However the poles of $\RHClr_{a,b}$ at $y_p=x_j$, $j=1,\dots,a$ are
compensated by the zeros of the prefactor:
\begin{equation}\label{Z-pref}
\fun(\bar\xi,\bar y)=\fun(\bar zq^{-2},\bar y)\fun(\bar xq^{-2},\bar y)=\fun^{-1}(\bar y,\bar z)\fun^{-1}(\bar y,\bar x)\,.
\end{equation}

It is easy to see that the r.h.s. $\tilde F^{(l,r)}_{a,b,p,n}$ has poles in the same points. Due to the product
$\fun(\bar y,\bar s_{\so})$ it has poles at $y_p=s_i$. The function $\Izerrl_p(\{\bar s_{\so}q^{-2},\bar\xi_{\so}\}|\bar y)$
has poles at $y_p=\xi_\ell$, however the poles at $y_p=s_iq^{-2}$ are compensated by the product $\fun(\bar y,\bar s_{\so})$.

Consider the residues of $F^{(l,r)}_{a,b,p,n}$ at $y_p=s_i$. Using the reduction property
\eqref{Rec-Z-triv1} we obtain (for shortness here and below we omit the arguments of $F^{(l,r)}_{a,b,p,n}$
and $\tilde F^{(l,r)}_{a,b,p,n}$ in the l.h.s. of equations):
\begin{multline}\label{F-res-ys}
F^{(l,r)}_{a,b,p,n}\Bigr|_{y_p\to s_i}= \fun(y_p,s_i)\fun(s_i,\bar s_i)\fun(\bar y_p,s_i)
\fun(\bar w,s_i)\bigl[\fun(s_i,\bar x)\fun(\bar xq^{-2},s_i)\bigr] \fun(\bar zq^{-2},s_i)\\
\times \fun(\bar\xi,\bar y_p) \RHClr_{a,b-1}(\bar t;\bar x|\bar s_i;\{\bar y_p,\bar w\})+\text{reg}.
\end{multline}
The terms in square brackets cancel each other, the terms in the second line give $F^{(l,r)}_{a,b-1,p-1,n}$:
\begin{equation}\label{F-res-ys1}
F^{(l,r)}_{a,b,p,n}\Bigr|_{y_p\to s_i}= \fun(y_p,s_i)\fun(s_i,\bar s_i)\fun(\bar y_p,s_i)
\fun(\bar w,s_i)\fun^{-1}(s_i,\bar z)
F^{(l,r)}_{a,b-1,p-1,n}(\bar t;\bar x|\bar s_i;\bar w;\bar y_p|\bar z)+\text{reg}.
\end{equation}

Consider now the residue of $\tilde F$ at $y_p=s_i$.
The pole occurs if and only if $s_i\in \bar s_{\so}$. Setting
$\bar s_{\so}=\{s_i,\bar s_0\}$ and using the property \eqref{K-red} of $\Izerrl$ we obtain
\begin{multline}\label{F-res-ys-r1}
\tilde F^{(l,r)}_{a,b,p,n}\Bigr|_{y_p\to s_i}=\sum (-q)^{\mp(k-1)}\Izerrl_{p-1}(\{\bar s_0q^{-2},\bar\xi_{\so}\}|\bar y_p)
\RHClr_{a,b-k}(\bar t;\bar x|\bar s_{\st};\{\bar w,\bar\xi_{\so}\})
\fun(s_i,\bar s_0)\fun(s_i,\bar s_{\st}) \\
\times \fun(\bar s_0,\bar s_{\st})\fun(\bar \xi_{\st},\bar \xi_{\so})\bigl[ \fun(y_p,s_i)\fun(\bar y_p, s_i)
\fun(\bar w, s_i)\fun^{-1}(s_i,\bar z)\bigr]\fun(\bar y_p,\bar s_0)\fun(\bar w,\bar s_0)\fun^{-1}(\bar s_0,\bar z)+\text{reg},
\end{multline}
where the sum is taking over partitions $\bar s_i\Rightarrow\{\bar s_0,\bar s_{\st}\}$ and $\bar \xi\Rightarrow\{\bar\xi_{\so},\bar\xi_{\st}\}$.
The terms in square brackets can be moved out of the sum. The product $\fun(s_i,\bar s_0)\fun(s_i,\bar s_{\st})$ combines into $\fun(s_i,\bar s_i)$ and also can be moved out of the sum. We arrive at
\begin{multline}\label{F-res-ys-r2}
\tilde F^{(l,r)}_{a,b,p,n}\Bigr|_{y_p\to s_i}= \fun(y_p,s_i)\fun(s_i,\bar s_i)\fun(\bar y_p,s_i)\fun(\bar w, s_i)\fun^{-1}(s_i,\bar z)
\sum (-q)^{\mp k_0}\Izerrl_{p-1}(\{\bar s_0q^{-2},\bar\xi_{\so}\}|\bar y_p)\\
\times  \RHClr_{a,b-1-k_0}(\bar t;\bar x|\bar s_{\st};\{\bar w,\bar\xi_{\so}\})\fun(\bar s_0,\bar s_{\st})\fun(\bar \xi_{\st},\bar \xi_{\so})
\fun(\bar y_p,\bar s_0)\fun(\bar w,\bar s_0) \fun^{-1}(\bar s_0,\bar z)+\text{reg},
\end{multline}
where $k_0=\#\bar s_0=k-1$. Evidently,  the sum over partitions in the r.h.s. of \eqref{F-res-ys-r2} gives
$\tilde F^{(l,r)}_{a,b-1,p-1,n}(\bar t;\bar x|\bar s_i;\bar w;\bar y_p|\bar z)$ and we obtain
\begin{equation}\label{F-res-ys-r3}
\tilde F^{(l,r)}_{a,b,p,n}\Bigr|_{y_p\to s_i}= \fun(y_p,s_i)\fun(s_i,\bar s_i)\fun(\bar y_p,s_i)\fun(\bar w, s_i)\fun^{-1}(s_i,\bar z)
\tilde F^{(l,r)}_{a,b-1,p-1,n}(\bar t;\bar x|\bar s_i;\bar w;\bar y_p|\bar z)+\text{reg}\,.
\end{equation}
Comparing \eqref{F-res-ys1} and \eqref{F-res-ys-r3} and taking into account \eqref{Ind-ass} we conclude that the
difference $F^{(l,r)}_{a,b,p,n}-\tilde F^{(l,r)}_{a,b,p,n}$ is a bounded function of $y_p$ as $y_p\to s_i$, $i=1,\dots, b$.
%
%

Consider now the residues of $F^{(l,r)}_{a,b,p,n}$ at $y_p=\xi_\ell$. We have
\begin{equation}\label{F-res-yxi}
F^{(l,r)}_{a,b,p,n}\Bigr|_{y_p\to \xi_\ell}
= \fun(\xi_\ell,y_p)\fun(\bar\xi_\ell,\xi_\ell)
\fun(\xi_\ell,\bar y_p)
\Bigl[\fun(\bar\xi_\ell,\bar y_p)\RHClr_{a,b}\bigl(\bar t;\bar x|\bar s;\bigl\{\bar w,\xi_\ell,\bar y_p\bigr\}\bigr)\Bigr]
+\text{reg}.
\end{equation}

Now one should distinguish between two cases: either $\xi_\ell\in \bar zq^{-2}$ or $\xi_\ell\in \bar xq^{-2}$.
Let $\xi_\ell=  z_jq^{-2}$. Then
the combination in the square brackets of \eqref{F-res-yxi} is just
$F^{(l,r)}_{a,b,p-1,n-1}(\bar t;\bar x|\bar s;\{\bar w,\xi_\ell\};\bar y_p|\bar z_j)$.
Thus, we obtain
\begin{equation}\label{F-res-yxi-z}
F^{(l,r)}_{a,b,p,n}\Bigr|_{y_p\to z_jq^{-2}}= \fun(\xi_\ell,y_p)\fun(\bar\xi_\ell,\xi_\ell)
\fun(\xi_\ell,\bar y_p) F^{(l,r)}_{a,b,p-1,n-1}(\bar t;\bar x|\bar s;\{\bar w,\xi_\ell\};\bar y_p|\bar z_j)+\text{reg}\,.
\end{equation}

Let now $\xi_\ell=  x_jq^{-2}$. In this case the prefactor $\fun(\bar\xi_\ell,\bar y_p)$ does not compensate the pole of $\RHClr_{a,b}$
at $y_p=x_j$. Therefore the combination in the squared brackets in \eqref{F-res-yxi} is not directly
$F^{(l,r)}_{a,b,p-1,n-1}$.

In order to overcome this problem we use
\eqref{dec-1} at $n=1$.  We have
\begin{multline}\label{dec-1-i}
\RHClr_{a,b}(\bar t;\{\bar x_j, x_j\}|\bar s;\{\bar w,\bar y_p,x_jq^{-2}\})=\sum_{i=1}^a \Izerlr_1(x_j|t_i)
\fun(t_i,\bar t_i)\fun(\bar x_j,t_i)\fun(\bar s,t_i)\\
\times \RHClr_{a-1,b}(\bar t_i;\bar x_j|\bar s;\{\bar w,t_iq^{-2},\bar y_p\})\,.
\end{multline}
Substituting \eqref{dec-1-i} into \eqref{F-res-yxi} we obtain
\begin{multline}\label{F-res-yxi-x0}
F^{(l,r)}_{a,b,p,n}\Bigr|_{y_p\to x_jq^{-2}}= \fun(\xi_\ell,y_p)\fun(\bar\xi_\ell,\xi_\ell)
\fun(\xi_\ell,\bar y_p)\sum_{i=1}^a \Izerlr_1(x_j|t_i) \fun(t_i,\bar t_i)\fun(\bar x_j,t_i)\fun(\bar s,t_i)\\
\times \fun(\bar\xi_\ell,\bar y_p)
\RHClr_{a-1,b}(\bar t_i;\bar x_j|\bar s;\{\bar w,t_iq^{-2},\bar y_p\})+\text{reg}\,.
\end{multline}
Now the combination in the second line of \eqref{F-res-yxi-x0} gives $F^{(l,r)}_{a-1,b,p-1,n}(\bar t_i;\bar x_j|\bar s;\{\bar w,t_iq^{-2}\};\bar y_p|\bar z)$, hence,
\begin{multline}\label{F-res-yxi-x}
F^{(l,r)}_{a,b,p,n}\Bigr|_{y_p\to x_jq^{-2}}= \fun(\xi_\ell,y_p)\fun(\bar\xi_\ell,\xi_\ell)
\fun(\xi_\ell,\bar y_p) \sum_{i=1}^a \Izerlr_1(x_j|t_i)\fun(t_i,\bar t_i)\fun(\bar x_j,t_i)\fun(\bar s,t_i)\\
\times
F^{(l,r)}_{a-1,b,p-1,n}(\bar t_i;\bar x_j|\bar s;\{\bar w,t_iq^{-2}\};\bar y_p|\bar z)+\text{reg}\,.
\end{multline}
Thus, we have reduced the residues of $F^{(l,r)}_{a,b,p,n}$ at $y_p=\xi_\ell$ to the functions $F^{(l,r)}_{a,b,p-1,n-1}$ or $F^{(l,r)}_{a-1,b,p-1,n}$.

Consider now the pole of $\tilde F^{(l,r)}_{a,b,p,n}$ at $y_p=\xi_\ell$. It  occurs if and only if $\xi_\ell\in \bar \xi_{\so}$. Setting
$\bar \xi_{\so}=\{\xi_\ell,\bar \xi_0\}$ and using property \eqref{K-Res} of $\Izerrl$ we obtain
\begin{multline}\label{F-res-yxi-r1}
\tilde F^{(l,r)}_{a,b,p,n}\Bigr|_{y_p\to \xi_\ell}=\sum (-q)^{\mp k}
\Izerrl_{p-1}(\{\bar s_{\so}q^{-2},\bar\xi_0\}|\bar y_p)
 \fun(\xi_\ell,y_p)\fun(\xi_\ell,\bar y_p)\fun(\bar s_{\so}q^{-2},\xi_\ell)
\fun(\bar\xi_0,\xi_\ell)\fun(\bar s_{\so},\bar s_{\st})\\
\times \RHClr_{a,b-k}(\bar t;\bar x|\bar s_{\st};\{\bar w,\bar\xi_0,\xi_\ell\}) \fun(\bar\xi_{\st},\bar\xi_0)
\fun(\bar\xi_{\st},\xi_\ell)\fun(\xi_\ell,\bar s_{\so})\fun(\bar y_p,\bar s_{\so})\fun(\bar w,\bar s_{\so})
\fun^{-1}(\bar s_{\so},\bar z)+\text{reg},
\end{multline}
where the sum is taking over partitions $\bar s\Rightarrow\{\bar s_{\so},\bar s_{\st}\}$ and
$\bar \xi_\ell\Rightarrow\{\bar\xi_{0},\bar\xi_{\st}\}$.
The terms $\fun(\bar s_{\so}q^{-2},\xi_\ell)$ and $\fun(\xi_\ell,\bar s_{\so})$ cancel each other.
The terms $ \fun(\xi_\ell,y_p)\fun(\xi_\ell,\bar y_p)$ can be moved out of the sum. The product $\fun(\bar\xi_0,\xi_\ell)\fun(\bar\xi_{\st},\xi_\ell)$ combines into $\fun(\bar \xi_\ell,\xi_\ell)$ and also can be moved out of the sum.  We arrive at
\begin{multline}\label{F-res-yxi-r2}
\tilde F^{(l,r)}_{a,b,p,n}\Bigr|_{y_p\to \xi_\ell}= \fun(\xi_\ell,y_p)\fun(\bar\xi_\ell,\xi_\ell)\fun(\xi_\ell,\bar y_p)\sum (-q)^{\mp k}
\Izerrl_{p-1}(\{\bar s_{\so}q^{-2},\bar\xi_0\}|\bar y_p)
\\
\times \RHClr_{a,b-k}(\bar t;\bar x|\bar s_{\st};\{\bar w,\bar\xi_0,\xi_\ell\}) \fun(\bar s_{\so},\bar s_{\st})\fun(\bar\xi_{\st},\bar\xi_0)
\fun(\bar y_p,\bar s_{\so})\fun(\bar w,\bar s_{\so})\fun^{-1}(\bar s_{\so},\bar z)+\text{reg}.
\end{multline}

Now  we set $\xi_\ell=z_jq^{-2}$. Then we simply rewrite
\begin{equation}\label{VS}
\fun(\bar w,\bar s_{\so})\fun^{-1}(\bar s_{\so},\bar z)=\Bigr[\fun(\bar w,\bar s_{\so})\fun(\xi_\ell,\bar s_{\so})\Bigl]\fun^{-1}(\bar s_{\so},\bar z_j),
\end{equation}
and we see that the sum over partitions in \eqref{F-res-yxi-r2} gives $\tilde F^{(l,r)}_{a,b,p-1,n-1}
(\bar t;\bar x|\bar s;\{\bar w,\xi_\ell\};\bar y_p|\bar z_j)$:
\begin{equation}\label{F-res-yxi-r2a}
\tilde F^{(l,r)}_{a,b,p,n}\Bigr|_{y_p\to z_jq^{-2}}= \fun(\xi_\ell,y_p)\fun(\bar\xi_\ell,\xi_\ell)\fun(\xi_\ell,\bar y_p)
\tilde F^{(l,r)}_{a,b,p-1,n-1}
(\bar t;\bar x|\bar s;\{\bar w,\xi_\ell\};\bar y_p|\bar z_j)
+\text{reg}.
\end{equation}

It remains to consider the case $\xi_\ell=x_jq^{-2}$. Due to \eqref{dec-1} we have
\begin{multline}\label{dec-2-i}
\RHClr_{a,b-k}(\bar t;\{\bar x_j, x_j\}|\bar s_{\st};\{\bar w,\bar\xi_0,x_jq^{-2}\})=\sum_{i=1}^a \Izerlr_1(x_j|t_i)
\fun(t_i,\bar t_i)\fun(\bar x_j,t_i)\fun(\bar s_{\st},t_i)\\
\times \RHClr_{a-1,b}(\bar t_i;\bar x_j|\bar s_{\st};\{\bar w,\bar\xi_0,t_iq^{-2}\}).
\end{multline}
Substituting \eqref{dec-2-i} into \eqref{F-res-yxi-r2} we  obtain
\begin{multline}\label{F-res-yxi-r3}
\tilde F^{(l,r)}_{a,b,p,n}\Bigr|_{y_p\to x_jq^{-2}}= \fun(\xi_\ell,y_p)\fun(\bar\xi_\ell,\xi_\ell)\fun(\xi_\ell,\bar y_p)
\sum_{i=1}^a \Izerlr_1(x_j|t_i)\fun(t_i,\bar t_i)\fun(\bar x_j,t_i)\\
\times \sum (-q)^{\mp k}\Izer_{p-1}(\{\bar s_{\so}q^{-2},\bar\xi_0\}|\bar y_p)
\RHClr_{a-1,b-k}(\bar t_i;\bar x_j|\bar s_{\st};\{\bar w,\bar\xi_0,t_iq^{-2}\}) \\
\times \fun(\bar s_{\so},\bar s_{\st})\fun(\bar\xi_{\st},\bar\xi_0)
\fun(\bar y_p,\bar s_{\so})\fun(\bar w,\bar s_{\so})\fun(\bar s_{\st},t_i)\fun^{-1}(\bar s_{\so},\bar z).
\end{multline}
Now we observe that
\begin{equation}\label{2-AVS}
\fun(\bar w,\bar s_{\so})\fun(\bar s_{\st},t_i)=\Bigl[\fun(\bar w,\bar s_{\so})\fun(t_iq^{-2},\bar s_{\so})\Bigr]\fun(\bar s,t_i).
\end{equation}
Substituting this into \eqref{F-res-yxi-r3} we arrive at
\begin{multline}\label{F-res-yxi-r4}
\tilde F^{(l,r)}_{a,b,p,n}(\bar t;\bar x|\bar s;\bar w;\bar y|\bar z)\Bigr|_{y_p\to \xi_\ell}= \fun(\xi_\ell,y_p)\fun(\bar\xi_\ell,\xi_\ell)\fun(\xi_\ell,\bar y_p)
\sum_{i=1}^a \Izerlr_1(x_j|t_i)\fun(t_i,\bar t_i)\fun(\bar x_j,t_i)\fun(\bar s,t_i)
\\
\times\sum (-q)^{\mp k}\Izerrl_{p-1}(\{\bar s_{\so}q^{-2},\bar\xi_0\}|\bar y_p)
\RHClr_{a-1,b-k}(\bar t_i;\bar x_j|\bar s_{\st};\{\bar w,\bar\xi_0,t_iq^{-2}\})
\fun(\bar s_{\so},\bar s_{\st})\fun(\bar\xi_{\st},\bar\xi_0)\\
\times \fun(\bar y_p,\bar s_{\so})\Bigl[\fun(\bar w,\bar s_{\so})\fun(t_iq^{-2},\bar s_{\so})\Bigr]\fun^{-1}(\bar s_{\so},\bar z)\,.
\end{multline}
Evidently the sum over partitions in \eqref{F-res-yxi-r4} gives $\tilde F^{(l,r)}_{a-1,b,p-1,n}(\bar t_i;\bar x_j|\bar s;\{\bar w,t_iq^{-2}\};\bar y_p|\bar z)$, therefore
\begin{multline}\label{F-res-yxi-r5}
\tilde F^{(l,r)}_{a,b,p,n}(\bar t;\bar x|\bar s;\bar w;\bar y|\bar z)\Bigr|_{y_p\to \xi_\ell}= \fun(\xi_\ell,y_p)\fun(\bar\xi_\ell,\xi_\ell)\fun(\xi_\ell,\bar y_p)
\\
\times \sum_{i=1}^a \Izerlr_1(x_j|t_i)\fun(t_i,\bar t_i)\fun(\bar x_j,t_i)\fun(\bar s,t_i)
F^{(l,r)}_{a-1,b,p-1,n}(\bar t_i;\bar x_j|\bar s;\{\bar w,t_iq^{-2}\};\bar y_p|\bar z)\,.
\end{multline}
Comparing \eqref{F-res-yxi-z} with \eqref{F-res-yxi-r2a} and \eqref{F-res-yxi-x} with \eqref{F-res-yxi-r5}, and taking into account
the induction assumption \eqref{Ind-ass}  we come to
conclusion that the difference $F^{(l,r)}_{a,b,p,n}-\tilde F^{(l,r)}_{a,b,p,n}$ is a bounded function of $y_p$ as $y_p\to\xi_\ell$,
$\ell=1,\dots a+n$.
%
%
Thus, we have proved that the function $F^{(l,r)}_{a,b,p,n}-\tilde F^{(l,r)}_{a,b,p,n}$  has no poles neither in the
points $y_p=s_i$ nor in $y_p=\xi_\ell$. Hence, this is a polynomial in $y_p$. Due to
\eqref{Klr-inf1} and  \eqref{Zlr-inf2} the polynomial
$F^{(r)}_{a,b,p,n}-\tilde F^{(r)}_{a,b,p,n}$ decreases as $y_p\to\infty$. Hence, $F^{(r)}_{a,b,p,n}-\tilde F^{(r)}_{a,b,p,n}=0$.
The case of the polynomial
$F^{(l)}_{a,b,p,n}-\tilde F^{(l)}_{a,b,p,n}$ is slightly more sophisticated. Due to
\eqref{Klr-inf2} and  \eqref{Zlr-inf1} we conclude that it is bounded as $y_p\to\infty$. Hence, it does not depend on $y_p$.
It follows from the representation \eqref{GF} that the highest coefficient $\RHCl_{a,b}(\bar t;\bar x|\bar s;\{\bar y,\bar w\})$
is proportional to the product of all $y_i$. Hence, the function $F^{(l)}_{a,b,p,n}$ vanishes at $y_p=0$. On the other hand
in the r.h.s. of \eqref{Sum-F} every term of the sum over partitions contains the right Izergin determinant $\Izerr_p(\{\bar s_{\so}q^{-2},\bar\xi_{\so}\}|\bar y)$. The latter also is proportional to the product of all $y_i$, and hence,  $\tilde F^{(l)}_{a,b,p,n}=0$  at $y_p=0$. Thus, we conclude that
$F^{(l)}_{a,b,p,n}-\tilde F^{(l)}_{a,b,p,n}=0$. \qed



\begin{thebibliography}{99}
%
\bibitem{FadST79} L. D. Faddeev, E. K. Sklyanin and L. A. Takhtajan, {\sl Quantum Inverse Problem. I},
 Theor. Math. Phys. {\bf 40} (1979) 688--706.
 %
 \bibitem{KulRes83}
P. P. Kulish, N. Yu. Reshetikhin,
{\sl Diagonalization of $GL(N)$ invariant transfer matrices and quantum $N$-wave system (Lee model)}, J.~Phys.~A:  {\bf 16} (1983) L591--L596.
%
\bibitem{BogIK93L}V. E. Korepin, N. M. Bogoliubov,
A. G. Izergin, {\sl Quantum Inverse Scattering Method and Correlation Functions}, Cambridge: Cambridge Univ.
Press, 1993.
%
\bibitem{FadLH96} L. D. Faddeev, in: Les Houches Lectures {\sl Quantum Symmetries}, eds A. Connes
et al, North Holland, (1998) 149.
%
\bibitem{Kor82} V. E. Korepin, {\sl Calculation of norms of Bethe wave functions}, Comm. Math. Phys. {\bf 86} (1982) 391--418.
%
\bibitem{IzeKor84}
A. G. Izergin,  V. E. Korepin,
{\sl The quantum inverse scattering method approach to correlation functions},
Comm. Math. Phys. {\bf 94} (1984), 67--92.
%
\bibitem{Kor84} V. E. Korepin, {\sl Correlation functions of the one-dimensional Bose gas in the repulsive case}, Comm. Math. Phys. {\bf 94}  (1984) 93--113.
    %
\bibitem{KitMaiT99}
N. Kitanine, J. M. Maillet, V. Terras, {\sl Form factors of the $XXZ$ Heisenberg spin-$1/2$ finite chain},
Nucl. Phys. B {\bf 554} (1999) 647--678, \texttt{arXiv:math-ph/9807020}.
%
\bibitem{MaiTer00} J. M. Maillet, V. Terras, {\sl On the quantum inverse scattering problem}, Nucl. Phys. B {\bf 575} (2000) 627--644, \texttt{hep-th/9911030}.
%
\bibitem{KitKMST07} N. Kitanine, K. Kozlowski, J. M. Maillet, N. A. Slavnov, V. Terras, {\sl
 On correlation  functions of integrable models associated to the six-vertex $R$-matrix}, J. Stat. Mech.
(2007) P01022, \texttt{arXiv:hep-th/0611142}.
%
\bibitem{KitMT00} N.~Kitanine, J.~M. Maillet, and V.~Terras,
{\sl  Correlation functions of the $XXZ$ Heisenberg spin-$1/2$ chain in a magnetic field},
Nucl. Phys. B {\bf 567} (2000) 554--582, \texttt{arXiv:math-ph/9907019}.
%
\bibitem{KitMST02} N. Kitanin, J. M. Maillet, N. A. Slavnov and V. Terras, {\sl Spin-spin correlation functions of the
$XXZ-1/2$ Heisenberg chain in a magnetic field}, Nucl. Phys. B {\bf 641} (2002) 487--518, \texttt{arXiv:hep-th/0201045}.
%
\bibitem{KitKMST09b} N. Kitanine, K.~K. Kozlowski, J. M. Maillet, N. A. Slavnov and V. Terras,
{\sl  Algebraic Bethe ansatz approach to the asymptotic behavior of correlation functions},
J. Stat. Mech. (2009) P04003, \texttt{ arXiv:0808.0227}.
%
\bibitem{GohKS04} F. G\"ohmann, A. Kl\"umper and A. Seel, {\sl  Integral representations for
correlation functions of the $XXZ$ chain at finite temperature},
J. Phys. A: Math. Gen. {\bf 37} (2004) 7625--7652, \texttt{arXiv:hep-th/0405089}.
%
%
\bibitem{GohKS05} F. G\"ohmann, A. Kl\"umper and A. Seel, {\sl  Integral representation of the density
matrix of the $XXZ$ chain at finite temperatures},
J. Phys. A: Math. Gen. {\bf 38} (2005) 1833--1842, \texttt{arXiv:cond-mat/0412062}.
%
\bibitem{SeeBGK07} A. Seel, T. Bhattacharyya, F. G\"ohmann, A. Kl\"umper,
{\sl  A note on the spin-$1/2$ $XXZ$ chain concerning its relation to the Bose gas}, J. Stat. Mech. (2007) P08030,
\texttt{arXiv:0705.3569}.
%
\bibitem{KitKMST11} N. Kitanine, K. Kozlowski, J. M. Maillet, N. A. Slavnov, V. Terras, {\sl
A form factor approach to the asymptotic behavior of correlation functions}, J. Stat. Mech. (2011) P12010,
\texttt{arXiv:hep-th/1110.0803}.
%
\bibitem{KitKMST12} N. Kitanine, K. Kozlowski, J. M. Maillet, N. A. Slavnov, V. Terras, {\sl
Form factor approach to dynamical correlation functions in critical models}, J. Stat. Mech. (2012) P09001,
\texttt{arXiv: 1206.2630}.
%
\bibitem{CauHM05} J.~S. Caux, J.~M. Maillet, {\sl  Computation of dynamical correlation functions of
Heisenberg chains in a field}, Phys. Rev. Lett. {\bf 95} (2005) 077201, \texttt{arXiv:cond-mat/0502365}.
%
\bibitem{PerSCHMWA06}
R.~G. Pereira, J. Sirker, J.~S. Caux, R. Hagemans, J.~M. Maillet, S.~R. White and I.~Affleck,
{\sl  The dynamical spin structure factor for the anisotropic spin-$1/2$ Heisenberg chain}, Phys. Rev. Lett. {\bf 96} (2006) 257202, \texttt{arXiv:cond-mat/0603681}.
%
\bibitem{PerSCHMWA07}
R.~G. Pereira, J. Sirker, J.~S. Caux, R. Hagemans, J.~M. Maillet, S.~R. White and I.~Affleck,
{\sl  Dynamical structure factor at small $q$ for the $XXZ$ spin-$1/2$ chain}, J. Stat. Mech. (2007) P08022,
\texttt{arXiv:0706.4327}.
%
\bibitem{CauCS07}
J.~S. Caux, P. Calabrese and N. A. Slavnov, {\sl  One-particle dynamical correlations in the one-dimensional Bose gas},
J. Stat. Mech. (2007) P01008, \texttt{arXiv:cond-mat/0611321}.
%
%
\bibitem{Sl} N. A. Slavnov, {\sl Calculation of scalar products of wave functions and form factors in the framework of the algebraic Bethe ansatz}, Theor. Math. Phys. {\bf 79}:2 (1989) 502--508.
    %
\bibitem{Res86}  N. Yu. Reshetikhin, {\sl Calculation of the norm of Bethe vectors in models with $SU(3)$-symmetry}, Zap. Nauchn. Sem. LOMI {\bf 150} (1986) 196--213;    J. Math. Sci. {\bf 46} (1989) 1694--1706 (Engl. transl.).
%
\bibitem{Whe12} M. Wheeler, {\sl Scalar products in generalized models with $SU(3)$-symmetry},
\texttt{arXiv:1204.2089}.
%
\bibitem{BelPRS12b} S. Belliard, S. Pakuliak, E. Ragoucy, N. A. Slavnov,
{\sl The algebraic Bethe ansatz for scalar products in $SU(3)$-invariant integrable
 models}, J. Stat. Mech. (2012) P10017, \texttt{arXiv:1207.0956}.
 %
\bibitem{BelPRS13a} S. Belliard, S. Pakuliak, E. Ragoucy, N. A. Slavnov,
{\sl Form factors in  $SU(3)$-invariant integrable models}, J. Stat. Mech.  (2013) P04033, \texttt{arXiv:1211.3968}.
 %
%
\bibitem{BelPRS12a} S. Belliard, S. Pakuliak, E. Ragoucy, N. A. Slavnov,
{\sl Highest coefficient of scalar products in $SU(3)$-invariant models}, J. Stat. Mech.  (2012) P09003, \texttt{arXiv:1206.4931}.
%
\bibitem{BelPRS13c}  S. Pakuliak, E. Ragoucy, N. A. Slavnov, {\sl Bethe vectors of quantum $\Uq{N}$-invariant integrable models},
\texttt{arXiv:1310.3253}.
%
 \bibitem{KulRes81}
P. P. Kulish, N. Yu. Reshetikhin,
{\sl Generalized Heisenberg ferromagnet and the Gross--Neveu model}, Zh. Eksp. Theor. Fiz.
{\bf 80} (1981) 214--228; Sov. Phys. JETP,  {\bf 53}:1 (1981)  108--114 (Engl. transl.)
%
 \bibitem{KulRes82}
P. P. Kulish, N. Yu. Reshetikhin,
{\sl GL(3)-invariant solutions of the Yang-Baxter equation and associated quantum systems}, Zap. Nauchn. Sem. POMI.
{\bf 120} (1982) 92--121; J. Sov. Math.,  {\bf 34}:5 (1982)  1948--1971 (Engl. transl.)
%
\bibitem{VT} V. Tarasov, A. Varchenko, {\sl Jackson inte\-gral re\-pre\-sen\-ta\-ti\-ons of so\-lu\-ti\-ons
of the quan\-tized Knizh\-nik--Za\-mo\-lod\-chi\-kov equation},  Algebra and Analysis, {\bf 6}:2 (1994) 90--137;
St. Petersburg Math. J. {\bf 6}:2 (1995) 275--313 (Engl. transl.), \texttt{arXiv:hep-th/9311040}.
%
\bibitem{KP-GLN} S. Khoroshkin, S. Pakuliak, {\sl A computation of an universal weight function for
the quantum affine algebra $\Uq{N}$.} {Journal of Mathematics of Kyoto University},
{\bf 48} n.2 (2008) 277--321.
%
\bibitem{EKhP} B. Enriquez, S. Khoroshkin, S. Pakuliak, {\sl Weight
functions and Drinfeld currents.}
{Comm. Math. Phys.} {\bf 276} (2007), 691--725.
%
\bibitem{OPS} A. Os'kin, S. Pakuliak, A. Silantyev, {\sl On the universal weight function
for the quantum affine algebra $\Uq{N}$.}
Algebra and Analysis  {\bf 21} n.4 (2009)   196--240.
%
\bibitem{PakRS13a} S. Pakuliak, E. Ragoucy, N. A. Slavnov,
{\sl Bethe vectors of quantum integrable models with $GL(3)$ trigonometric $R$-matrix},
SIGMA \textbf{9} (2013) 058, \texttt{arXiv:1304.7602}.
%
%
\bibitem{BelDri82} A. A. Belavin and V. G. Drinfel'd, {\sl Solutions of the classical Yang-Baxter equation for simple Lie algebras},
    Functional Analysis and Its Applications, (1982) \textbf{16}:3 159--180.
%
\bibitem{Ize87} A. G. Izergin, {\sl Partition function of the six-vertex model in a finite volume},
Dokl. Akad. Nauk SSSR {\bf 297} (1987) 331--333;
Sov. Phys. Dokl. {\bf 32} (1987) 878--879 (Engl. transl.).
%
\bibitem{BPR} S. Belliard, S. Pakuliak, E. Ragoucy, {\sl Universal Bethe Ansatz and Scalar Products of Bethe Vectors }, SIGMA {\bf 6} (2010) 94, \texttt{arXiv:1012.1455}.
%
%
\end{thebibliography}
\end{document}